# FAST MULTIPOLE BOUNDARY ELEMENT METHOD FOR THREE DIMENSIONAL ELECTROMAGNETIC SCATTERING PROBLEM

S. B. WANG*

*Department of Physics and William Mong Institute of Nano Science and Technology,*
*The Hong Kong University of Science and Technology,*
*Clear Water Bay, Kowloon, Hong Kong, P. R. China*
*wangsb@ust.hk*

H. H. ZHENG

*Department of Physics ,*
*The Hong Kong University of Science and Technology,*
*Clear Water Bay, Kowloon, Hong Kong, P. R. China*
*zhenghh@ust.hk*

J. J. XIAO

*Department of Electronic and Information Engineering,*
*Shenzhen Graduate School, Harbin Institute of Technology,*
*Shenzhen 518055, P. R. China*
*xiaoxun@gmail.com*

Z. F. LIN

*Key Laboratory of Micro and Nano Photonic Structures (Ministry of Education),*
*Fudan University,*
*Shanghai 200433, P. R. China*
*phlin@fudan.edu.cn*

C. T. CHAN

*Department of Physics and William Mong Institute of Nano Science and Technology,*
*The Hong Kong University of Science and Technology,*
*Clear Water Bay, Kowloon, Hong Kong, P. R. China*
*phchan@ust.hk*

We developed a fast numerical algorithm for solving the three dimensional vectorial Helmholtz equation that arises in electromagnetic scattering problems. The algorithm is based on electric field integral equations and is essentially a boundary element method. Nyström's quadrature rule with a triangular grid is employed to linearize the integral equations, which are then solved by using a right-preconditioned iterative method. We apply the fast multipole technique to accelerate the matrix-vector multiplications in the iterations. We demonstrate the broad applications and accuracy of this method with practical examples including dielectric, plasmonic and metallic objects. We then apply the method to investigate the plasmonic properties of a silver torus and a silver split-ring resonator under the incidence of an electromagnetic plane wave. We show the silver torus can be used as a trapping tool to bind small dielectric or metallic particles.

## 1. Introduction

In recent years the fast development in the field of metamaterials and plasmonics [Maier, 2007; Zouhdi *et al*., 2009] provides inspirations for new nano-photonic devices. Various metallic nanostructures such as nanospheres [Liu *et al*., 2011], nanowires [Kottmann and Martin, 2001], nanorods [Huang *et al*., 2009], nanopatches [Liu *et al*., 2011; Wang *et al*., 2011], as well as their ordered or disordered aggregates [Vallecchi *et al*., 2011], have been investigated. These structured metallic systems can interact strongly with light due to the resonance excitation of the surface plasmon modes, which give rise to an enhancement of the local fields [Lindberg *et al*., 2004]. The sensitive dependence of the plasmon resonance on the geometrical shapes of the structures provides broad tunability for its applications. However, for the structures that do not have a simple shape (e.g. sphere) one has to resort to some numerical methods to simulate or predict the electromagnetic responses of these complex systems. These numerical methods are broadly categorized into two major types: methods based on differential equations and methods based on integral equations [Myroshnychenko *et al*., 2008; Myroshnychenko *et al*., 2008]. Methods of the first type such as the finite-difference time-domain (FDTD) method [Taflove and Hagness, 2005] surely can provide fairly accurate computations. But they do have certain disadvantages, especially when one needs to examine the further consequences of the shape effects on local field distribution, for instance, the modeling of optical micromanipulations [Chen *et al*., 2011; Dholakia and Reece, 2006; Grier, 2003], optical binding [Grzegorczyk *et al*., 2006; Ng *et al*., 2005] as well as near-field optical forces [Chaumet *et al*., 2002]. The consuming of the computing resources and time may become a problem in such studies. On the contrary, numerical treatment of electromagnetic problems via integral equations is very attractive in addressing these problems. The reason lies on the fact that the integral-equation methods can convert a homogenous volumetric problem into a boundary problem, thus reducing the dimensionalities by one [Kern and Martin, 2009; Wrobel and Aliabadi, 2005; Xiao and Chan, 2008; Xiao *et al*., 2009]. One of such methods is the boundary element method (BEM). In addition to the reduced unknowns and its extreme flexibility with respect to the geometry, the approach does not suffer from the staggered field value problem due to the discretization of the computation domain. This makes the BEM more convenient at addressing certain post-processing problems such as those in optical force calculation [Wang *et al*., 2011] using the Maxwell tensor approach which requires information of both the electric and magnetic fields at the same grid points.

The BEM is rooted in the Green's theorem [Jackson, 1999] which enables one to construct local fields value in all the computational domains with just boundary fields and their derivatives. To implement this method, the boundary integral equations are first formulated with the homogeneous Green's function and the corresponding boundary conditions. A quadrature rule is then applied to linearize the obtained equations, making them numerically solvable by either a direct method (e.g. Gauss elimination [Atkinson, 1989]) or an iterative method (e.g. fGMRES [Fraysse *et al.*, 2008; Saad, 1993; Saad, 2003]).

Although it has reduced the dimensionalities by one, the conventional BEM is not necessarily superior to a domain method in three-dimensional (3D) cases. This is because the BEM produces a dense asymmetric matrix. Hence the computational complexity of the BEM is $O(N^2) = O(n^4)$ (*N* is the total number of elements, *n* is the number of nodes along each dimension and $N \sim n^2$ as the elements are confined on a 2D surface) when integrated with an iterative solver. However, for the finite element method (FEM), the computational complexity is $O(N) = O(n^3)$ since it produces sparse matrix [Zienkiewicz *et al.*, 2005]. A remedy is to combine the fast multipole method (FMM) [Chew *et al.*, 2001; Engheta *et al.*, 1992] with the BEM, which accelerates the matrix-vector multiplications in the iterations. The improved method benefits from its high accuracy, fast solution as well as less memory and time consumed. A review of the general principle of this method can be found in the literature (see for example the review by Nishimura [2002]).

We note even though FMM has already been integrated with integral-equation based methods to solve electromagnetic problems [Engheta *et al.*, 1992; Liu, 2009; Song *et al.*, 1997; Wu *et al.*, 2012], most of them use the method of moments (MOM) to solve the integral equations. In 3D electrodynamics, the BEM has not been implemented with the Nyström's quadrature rule. The main difference is that the MOM formulates the integral equations with electric and magnetic currents, which is expanded in certain basis functions. Hence the unknowns of the MOM are the weights of the basis functions. In the BEM, the integral equations are formulated with electric/magnetic field directly and discretized by convergent quadrature formulas. The unknowns are the fields' value and their derivatives. The advantage of the BEM is that it is generally easier to apply higher order quadrature method than to employ higher order basis functions [Engheta *et al.*, 1992]. Besides, the matrix filling is more efficient with Nyström's quadrature method than with the moment method.

We present in this paper the fast multipole boundary element method (FMBEM) developed for solving 3D electromagnetic scattering problems. The method is formulated with electric integral equations and employs the Nyström's quadrature rule to discretize the obtained equations. The iterative fGMRES method with a block-diagonal pre-conditioner is utilized to solve the linearized equations. We show by complexity study that the FMBEM is an $O(n_{\text{iter}} N \log N)$ ($n_{\text{iter}}$ is the total number of iterations) method. Hence it is a practical and efficient solver for large scale problems. We also show by practical examples that the developed method can address varieties of problems involving dielectric, plasmonic or metallic systems. The outline of this paper is as follows. In Section 2 we present the detailed formalisms and the numerical schemes used in the FMBEM. In Section 3, we demonstrate the accuracy and the efficiency of this method by comparing its results with Mie scattering theory or experimental data. In Section 4, we apply the developed method to investigate the plasmonic properties of a torus and a split-ring resonator (SRR). We also studied the optical forces induced in a sphere-torus structure, which shows that a silver torus can be used as a trapping tool. Some concluding remarks are then made in Section 5.

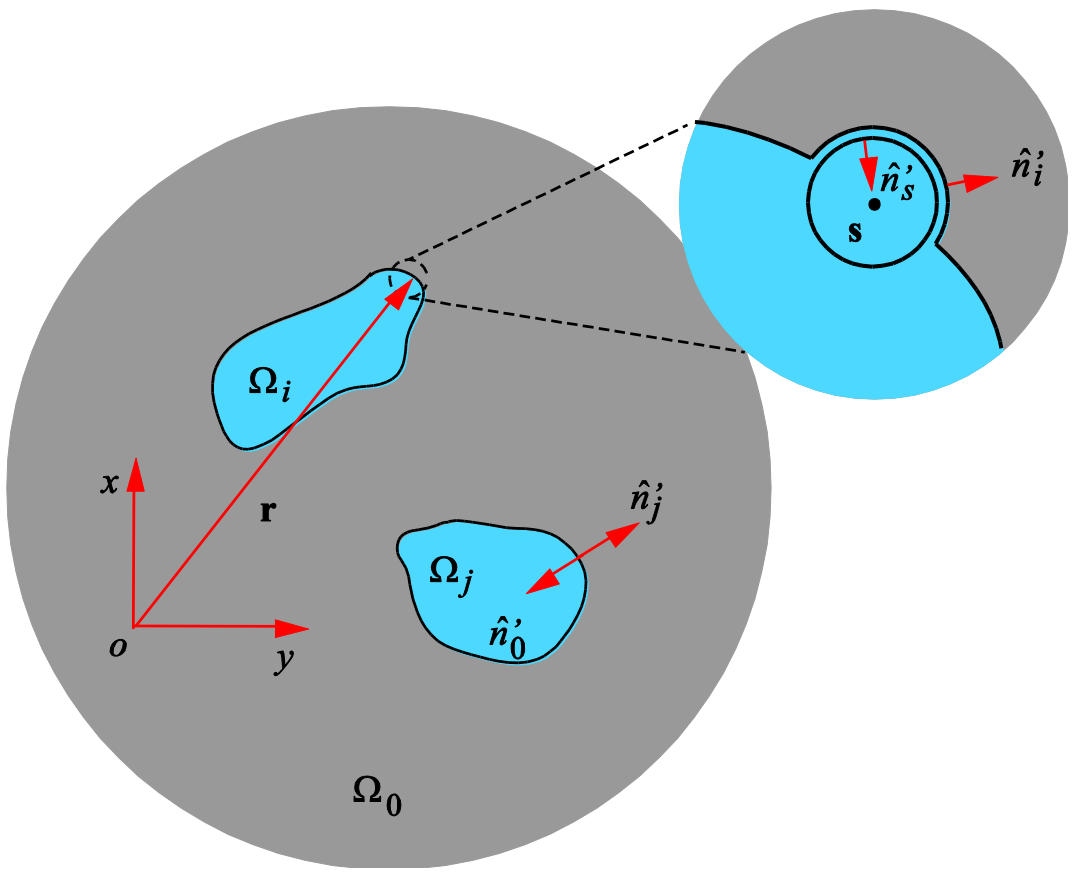

Fig. 1. A pictorial representation of a typical scattering system. $\Omega_0$ is the open space background where the incident wave lies in; $\Omega_i$, $\Omega_j$ are two arbitrary scatters; $\hat{n}_0'$, $\hat{n}_i'$, $\hat{n}_j'$ are the outward unit normal vectors on the boundary of the domains $\Omega_0$, $\Omega_i$, $\Omega_j$. An infinitesimal sphere is extracted out to solve the singular integral problem. The unit normal vector $\hat{n}_s'$ of this infinitesimal sphere has an inward direction.

## 2. Formulations and numerical implementation

### 2.1. *Boundary integral equations*

Let us consider an open system consisting of $J$ arbitrarily-shaped 3D scatters that are homogeneous and isotropic (Fig. 1). We assume that the electromagnetic fields are time-harmonic, with a conventional time factor $e^{-i\omega t}$ throughout this paper. In the frequency domain, the electric fields in different regions are determined by the vectorial Helmholtz equation

$$\nabla \times \left[ \nabla \times \mathbf{E}_i(\mathbf{r}) \right] - k_i^2 \mathbf{E}_i(\mathbf{r}) = i\omega\mu_i \mathbf{j}(\mathbf{r}),\ \mathbf{r} \in \Omega_i, i = 0, 1, ..., J\ , \tag{2.1}$$

where $i = 0$ denotes the background domain; $k_i^2 = \varepsilon_i \mu_i \omega^2 / c^2$ with $k_i$ being the wavenumber of the electromagnetic wave in the domain $\Omega_i$ and $j$ denotes the source current density; $c$ is the speed of light in vacuum; $\varepsilon_i$ and $\mu_i$ denote the relative permittivity and permeability of the medium; $\omega$ is the harmonic angular frequency. The corresponding dyadic Green's function satisfies

$$\nabla \times \left[ \nabla \times \overline{\mathbf{G}}_i(\mathbf{r}', \mathbf{r}) \right] - k_i^2 \overline{\mathbf{G}}_i(\mathbf{r}', \mathbf{r}) = \delta(\mathbf{r}' - \mathbf{r}) \overline{\mathbf{I}}\ , \tag{2.2}$$

where $\overline{\mathbf{I}}$ is a $3 \times 3$ unit tensor with elements $\overline{\mathbf{I}}_{ij} = \delta_{ij}$ and $\delta_{ij}$ is the Kronecker delta. The dyadic Green's function is expressed as [Tsang *et al*., 2000]

$$\overline{\mathbf{G}}_i(\mathbf{r}',\mathbf{r}) = g_i(\mathbf{r}',\mathbf{r})\overline{\mathbf{I}} + \frac{\nabla\nabla}{k_i^2} g_i(\mathbf{r}',\mathbf{r}) . \tag{2.3}$$

Here $g_i(\mathbf{r}',\mathbf{r})$ is the 3D scalar Green's function that takes the form of

$$g_i(\mathbf{r}',\mathbf{r}) = \frac{e^{ik_i|\mathbf{r}'-\mathbf{r}|}}{4\pi\,|\mathbf{r}'-\mathbf{r}|} . \tag{2.4}$$

We multiply Eq. (2.1) with $\overline{\mathbf{G}}_i(\mathbf{r}',\mathbf{r})$ and subtract the resulting equation from Eq. (2.2) multiplied by $\mathbf{E}_i(\mathbf{r})$. Then we take volume integrals on both sides of the resulting equation with respect to each domain and apply Green's second identity to transform the left-hand-side part into a surface integral. The electric field integral equations (EFIEs) can be obtained after some tedious manipulations (see the Appendix A):

$$\begin{aligned}
&\overset{\mathrm{CPV}}{\oint_{\partial\Omega_i}} \mathrm{d}A' \begin{Bmatrix} \left[\hat{n}_i'\nabla' g_i(\mathbf{s}',\mathbf{s}) - \hat{n}_i'\cdot\nabla' g_i(\mathbf{s}',\mathbf{s})\overline{\mathbf{I}} - \nabla' g_i(\mathbf{s}',\mathbf{s})\hat{n}_i'\right]\cdot\mathbf{E}_i(\mathbf{s}') \\ -g_i(\mathbf{s}',\mathbf{s})\overline{\mathbf{I}}\cdot\left[\hat{n}_i'\times(\nabla'\times\mathbf{E}_i(\mathbf{s}'))\right] \end{Bmatrix} \\
&= \frac{1}{2}\mathbf{E}_i(\mathbf{s}) - \mathbf{E}_i^{\mathrm{inc}}(\mathbf{s}),\ i = 0,1,\ldots,J.
\end{aligned} \tag{2.5}$$

Here "CPV" indicates the integral is a Cauchy-Principle-Value integral. $\mathbf{E}_i^{\mathrm{inc}}(\mathbf{s})$ is the incident electric field in the domain $\Omega_i$; s' denotes the source point and s denotes the evaluation point (we have replaced $\mathbf{r}'$, $\mathbf{r}$ with $\mathbf{s}'$, $\mathbf{s}$ to indicate the points locate on the boundary); $\hat{n}_i'$ is the unit normal vector at the source point that points outside of the domain $\Omega_i$; $\partial\Omega_i$ refers to the boundary of the domain $\Omega_i$. The partial derivative $\nabla'$ is taken with respect to the source point.

The boundary field values inside and outside a domain are related by the following boundary conditions

$$\mathbf{E}_i(\mathbf{s}) = \left(\overline{\mathbf{I}} + \frac{\epsilon_0 - \epsilon_i}{\epsilon_i}\hat{n}_i'\hat{n}_i'\right)\cdot\mathbf{E}_0(\mathbf{s}) , \tag{2.6}$$

$$\hat{n}_i'\times(\nabla\times\mathbf{E}_i(\mathbf{s})) = \frac{\mu_i}{\mu_0}\hat{n}_i'\times(\nabla\times\mathbf{E}_0(\mathbf{s})),\ i = 1,2,\ldots,J. \tag{2.7}$$

Equation (2.6) represents the continuity of the tangential component of the electric field (i.e., $\hat{n}_i'\times(\mathbf{E}_i - \mathbf{E}_0) = 0$) and the normal component of the electric displacement field (i.e., $\hat{n}_i'\cdot(\varepsilon_i\mathbf{E}_i - \varepsilon_0\mathbf{E}_0) = 0$). Eq. (2.7) represents the continuity of the tangential components of the magnetic field (i.e., $\hat{n}_i'\times(\mathbf{H}_i - \mathbf{H}_0) = 0$).

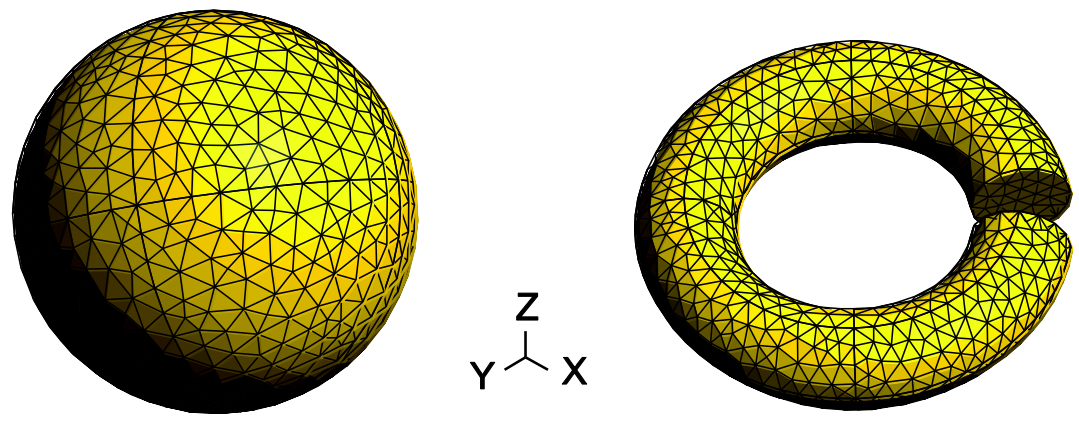

Fig. 2. Planar triangle grids for a sphere and a split ring resonator.

## 2.2. *Numerical implementation and the fast multipole technique*

In the numerical implementation, Eq. (2.5) is discretized by a grid consisting of $N$ planar triangular elements [Geuzaine and Remacle, 2009] (see Fig. 2) with $N=\sum_{i=1}^{J} u_i$, where $u_i$ is the total number of elements contained in the scatter $\Omega_i$ and $J$ is the total number of scatters. Hence, for the background domain we have $u_0 = N$. The integral of Eq. (2.5) can be approximated by using the Nyström's quadrature rule:

$$\oint_{\partial\Omega_i}^{\text{CPV}} \mathrm{d}A' f(\mathbf{s}',\mathbf{s}) \approx \sum_{j=1}^{u_i} w(\mathbf{s}'_j) f(\mathbf{s}'_j,\mathbf{s}), \tag{2.8}$$

where $f(\mathbf{s}',\mathbf{s})$ represents the integrand in Eq. (2.5) and $w(\mathbf{s}'_j)$ is the weight of the $j_{\text{th}}$ element. In the simplest case, we take $f(\mathbf{s}',\mathbf{s})$ to be a constant over a triangular element and $\mathbf{s}'_j$ is taken to be the element's mass center. $w(\mathbf{s}'_j)$ is taken to be the area of the element. Consequently, the problem is reduced to solving $6N$ linear algebraic equations that can be written in a compact form as

$$\begin{bmatrix} \bar{\mathbf{\Gamma}} & \bar{\mathbf{\Lambda}} \end{bmatrix} \begin{bmatrix} \mathbf{\Phi} \\ \mathbf{\Psi} \end{bmatrix} = \mathbf{\Phi}^{\text{inc}}, \tag{2.9}$$

where

$$\bar{\mathbf{\Gamma}} = \begin{bmatrix} \bar{\mathbf{\Gamma}}_0 & & & \\ & \bar{\mathbf{\Gamma}}_1 & & \\ & & \ddots & \\ & & & \bar{\mathbf{\Gamma}}_J \end{bmatrix}, \bar{\mathbf{\Lambda}} = \begin{bmatrix} \bar{\mathbf{\Lambda}}_0 & & & \\ & \bar{\mathbf{\Lambda}}_1 & & \\ & & \ddots & \\ & & & \bar{\mathbf{\Lambda}}_J \end{bmatrix},$$
$$\mathbf{\Phi} = \begin{bmatrix} \mathbf{\Phi}_1 \\ \mathbf{\Phi}_2 \\ \vdots \\ \mathbf{\Phi}_J \end{bmatrix}, \mathbf{\Psi} = \begin{bmatrix} \mathbf{\Psi}_1 \\ \mathbf{\Psi}_2 \\ \vdots \\ \mathbf{\Psi}_J \end{bmatrix}, \mathbf{\Phi}^{\text{inc}} = \begin{bmatrix} \mathbf{\Phi}_0^{\text{inc}} \\ \mathbf{\Phi}_1^{\text{inc}} \\ \vdots \\ \mathbf{\Phi}_J^{\text{inc}} \end{bmatrix} \tag{2.10}$$

and

$$\bar{\mathbf{\Gamma}}_i = \begin{bmatrix} \bar{\mathbf{\Gamma}}_{i,11} & \bar{\mathbf{\Gamma}}_{i,12} & \cdots & \bar{\mathbf{\Gamma}}_{i,1u_i} \\ \bar{\mathbf{\Gamma}}_{i,21} & \bar{\mathbf{\Gamma}}_{i,22} & \cdots & \bar{\mathbf{\Gamma}}_{i,2u_i} \\ \vdots & \vdots & \ddots & \vdots \\ \bar{\mathbf{\Gamma}}_{i,u_i1} & \bar{\mathbf{\Gamma}}_{i,u_i2} & \cdots & \bar{\mathbf{\Gamma}}_{i,u_iu_i} \end{bmatrix},$$

$$\bar{\mathbf{\Lambda}}_i = \begin{bmatrix} \bar{\mathbf{\Lambda}}_{i,11} & \bar{\mathbf{\Lambda}}_{i,12} & \cdots & \bar{\mathbf{\Lambda}}_{i,1u_i} \\ \bar{\mathbf{\Lambda}}_{i,21} & \bar{\mathbf{\Lambda}}_{i,22} & \cdots & \bar{\mathbf{\Lambda}}_{i,2u_i} \\ \vdots & \vdots & \ddots & \vdots \\ \bar{\mathbf{\Lambda}}_{i,u_i1} & \bar{\mathbf{\Lambda}}_{i,u_i2} & \cdots & \bar{\mathbf{\Lambda}}_{i,u_iu_i} \end{bmatrix}, \tag{2.11}$$

$$\mathbf{\Phi}_i = \begin{bmatrix} \mathbf{\Phi}_{i,1} \\ \mathbf{\Phi}_{i,2} \\ \vdots \\ \mathbf{\Phi}_{i,u_i} \end{bmatrix}, \ \mathbf{\Psi}_i = \begin{bmatrix} \mathbf{\Psi}_{i,1} \\ \mathbf{\Psi}_{i,2} \\ \vdots \\ \mathbf{\Psi}_{i,u_i} \end{bmatrix}, \ \mathbf{\Phi}_i^{\text{inc}} = \begin{bmatrix} \mathbf{\Phi}_{i,1}^{\text{inc}} \\ \mathbf{\Phi}_{i,2}^{\text{inc}} \\ \vdots \\ \mathbf{\Phi}_{i,u_i}^{\text{inc}} \end{bmatrix}.$$

The matrix elements in Eq. (2.11) take the following forms:

$$\begin{aligned} \bar{\mathbf{\Gamma}}_{i,\alpha\beta} &= \frac{1}{2}\delta_{\alpha\beta}\bar{\mathbf{I}} - \int_{\mathbf{s}_\beta \in \partial\Omega_i} \mathrm{d}A' \begin{bmatrix} \hat{n}_i' \nabla' g_i(\mathbf{s}_\beta, \mathbf{s}_\alpha) - \hat{n}_i' \cdot \nabla' g_i(\mathbf{s}_\beta, \mathbf{s}_\alpha)\bar{\mathbf{I}} \\ -\nabla' g_i(\mathbf{s}_\beta, \mathbf{s}_\alpha)\hat{n}_i' \end{bmatrix} \cdot \bar{\mathbf{T}}_i, \\ \bar{\mathbf{\Lambda}}_{i,\alpha\beta} &= \gamma_i \int_{\mathbf{s}_\beta \in \partial\Omega_i} \mathrm{d}A' g_i(\mathbf{s}_\beta, \mathbf{s}_\alpha)\bar{\mathbf{I}}, \\ \mathbf{\Phi}_{i,\beta} &= \mathbf{E}_0(\mathbf{s}_\beta), \\ \mathbf{\Psi}_{i,\beta} &= \hat{n}_i' \times (\nabla' \times \mathbf{E}_0(\mathbf{s}_\beta)), \\ \mathbf{\Phi}_{i,\alpha}^{\text{inc}} &= \mathbf{E}_i^{\text{inc}}(\mathbf{s}_\alpha), \end{aligned} \tag{2.12}$$

where we have defined

$$\bar{\mathbf{T}}_i = \begin{cases} \bar{\mathbf{I}} & i = 0; \\ \bar{\mathbf{I}} + \dfrac{\epsilon_0 - \epsilon_i}{\epsilon_i}\hat{n}_i'\hat{n}_i' & i = 1,2,\ldots,J. \end{cases}, \ \gamma_i = \begin{cases} 1 & i = 0; \\ \dfrac{\mu_i}{\mu_0} & i = 1,2,\ldots,J. \end{cases}. \tag{2.13}$$

Here $\mathbf{s}_\alpha$ and $\mathbf{s}_\beta$ denote the mass center of the element $\alpha$ and $\beta$, respectively and $\delta_{\alpha\beta}$ is the Kronecker delta. The singular integral in Eq. (2.12) can be handled analytically (see the Appendix B). To solve Eq. (2.9), we employ an iterative fGMRES method here. We note that during the iterations, the most time-consuming step is the matrix-vector multiplication corresponding to the left-hand side of Eq. (2.9). A direct multiplication has $O(N^2)$ complexity which becomes the bottleneck for large scale systems (where $N$ is a very big number), especially when the iteration has poor convergence property. One way to remedy this is to apply FMM that can substantially reduce the operations.

In the BEM, each triangle element is a source and a receiver at the same time. From a physical point of view, operators $\bar{\mathbf{\Gamma}}$ and $\bar{\mathbf{\Lambda}}$ in Eq. (2.9) pass the interactions from the sources $\mathbf{\Phi}$ and $\mathbf{\Psi}$ to the receivers through the Green's function. Both the information of the sources and the receivers is needed for calculating their interactions. Hence, we need to do the one-to-one interaction-passing work for each group of source and receiver. However, if the receivers and the sources can be separated so that the interactions can be calculated without knowing the information of the receivers, we can group the interactions and then spread them to the receivers. This will reduce the total operations substantially and is the essential idea of the FMBEM.

The first step to separate the sources from the receivers is to expand the Green's function $g(\mathbf{r}',\mathbf{r})$ at an arbitrary point $\mathbf{r}^{\star}$ with a multipole basis as

$$\begin{aligned} g(\mathbf{r}',\mathbf{r}) &= \frac{e^{ik|\mathbf{r}'-\mathbf{r}|}}{4\pi|\mathbf{r}'-\mathbf{r}|} \\ &= ik\sum_{n=0}^{\infty}\sum_{m=-n}^{n} j_n(kd)h_n^{(1)}(kD)Y_n^{-m}(\theta_d,\varphi_d)Y_n^{m}(\theta_D,\varphi_D), \end{aligned} \tag{2.14}$$

where $\mathbf{d}=\mathbf{r}'-\mathbf{r}^{\star}, \mathbf{D}=\mathbf{r}-\mathbf{r}^{\star}$ ( $|\mathbf{d}|<|\mathbf{D}|$ ) and $j_n(kd)$ is the spherical Bessel function. $h_n^{(1)}(kD)$ is the spherical Hankel function of the first kind (we will omit the superscript in the following discussion). $Y_n^m(\theta,\varphi)$ is the spherical harmonics defined as

$$\begin{aligned} &Y_n^m(\theta,\varphi) = (-1)^m\sqrt{\frac{2n+1}{4\pi}\frac{(n-|m|)!}{(n+|m|)!}}P_n^{|m|}(\cos\theta)e^{im\varphi}, \\ &n=0,1,2,\ldots,\ m=-n,-n+1,\ldots,n-1,n, \end{aligned} \tag{2.15}$$

where $P_n^{|m|}$ is the associated Legendre function. $\theta$ and $\varphi$ are the polar angle and azimuthal angle in the spherical coordinates, respectively. We have the relationships $\mathbf{d}=d(\sin\theta_d\cos\varphi_d,\sin\theta_d\sin\varphi_d,\cos\theta_d)$ and $\mathbf{D}=D(\sin\theta_D\cos\varphi_D,\sin\theta_D\sin\varphi_D,\cos\theta_D)$ . Let us define

$$\begin{aligned} R_n^m(\mathbf{r}-\mathbf{r}^{\star}) &= j_n(k|\mathbf{r}-\mathbf{r}^{\star}|)Y_n^m(\theta,\varphi), \\ S_n^m(\mathbf{r}-\mathbf{r}^{\star}) &= h_n(k|\mathbf{r}-\mathbf{r}^{\star}|)Y_n^m(\theta,\varphi). \end{aligned} \tag{2.16}$$

$\{R_n^m(\mathbf{r}-\mathbf{r}^{\star})\}$ and $\{S_n^m(\mathbf{r}-\mathbf{r}^{\star})\}$ are then referred to as the regular bases and the singular bases with respect to the different properties of spherical Bessel function and Hankel function at $\mathbf{r}=\mathbf{r}^{\star}$ . With Eq. (2.16) the Green's function expansions can be rewritten as

$$g(\mathbf{r}',\mathbf{r}) = \begin{cases} ik\displaystyle\sum_{n=0}^{\infty}\sum_{m=-n}^{n} R_n^{-m}(\mathbf{r}'-\mathbf{r}^{\star})S_n^m(\mathbf{r}-\mathbf{r}^{\star}), & \left|\mathbf{r}'-\mathbf{r}^{\star}\right| < \left|\mathbf{r}-\mathbf{r}^{\star}\right|; \\ ik\displaystyle\sum_{n=0}^{\infty}\sum_{m=-n}^{n} S_n^{-m}(\mathbf{r}'-\mathbf{r}^{\star})R_n^m(\mathbf{r}-\mathbf{r}^{\star}), & \left|\mathbf{r}'-\mathbf{r}^{\star}\right| > \left|\mathbf{r}-\mathbf{r}^{\star}\right|. \end{cases} \tag{2.17}$$

The upper-branch of Eq. (2.17) is the far-field expansion while the lower-branch is the near-field expansion. In the numerical implementations, the summations in the above equations are truncated at $n=p-1$, where $p$ is referred to as the truncation number. By applying the Green's function expansions the source ( $\mathbf{r}'$ ) and the receiver ( $\mathbf{r}$ ) are decomposed into separate terms. Hence we can process either of them without referring to the other one.

The FMBEM is built on a hierarchical octree data structure (see the Appendix C for the detail). The computation domain is first enclosed in a cubic box on level $l=0$. Then the box is subdivided into $2^3$ "children" boxes on level $l=1$ with each box contains a few number of the elements. This process is continued until on level $l=l_{\max}$ there are only several elements contained in each box. A global index $(l,q)$ is used to label the $q_{\text{th}}$ box on level *l*. With these definitions, the matrix-vector multiplication (interaction-passing) can be done by the following four steps.

*Step 1. Collect initial information*

On the finest level $l=l_{\max}$, the interaction of each element is passed to its box . This is done by expanding the Green's function at the far-field with respect to the box's center $\mathbf{r}^{\star}$ . Substitute Eq. (2.17) into Eq. (2.12), the expansion coefficient vector (the interaction) produced by the element $\mathbf{s}_{\beta}$ can be written as

$$\mathbf{C}_i^{\mathbf{s}_\beta} = ik_i \int_{\mathbf{s}_\beta \in \partial\Omega_i} \mathrm{d}A' \left\{ \begin{array}{l} -\left[ \begin{array}{l} \hat{n}_i' \nabla' R_i(\mathbf{s}_\beta, \mathbf{r}^\star) - \hat{n}_i' \cdot \nabla' R_i(\mathbf{s}_\beta, \mathbf{r}^\star) \overline{\mathbf{I}} \\ -\nabla' R_i(\mathbf{s}_\beta, \mathbf{r}^\star) \hat{n}_i' \end{array} \right] \cdot \overline{\mathbf{T}}_i \cdot \mathbf{E}_0(\mathbf{s}_\beta) \\ + \gamma_i R_i(\mathbf{s}_\beta, \mathbf{r}^\star) \left[ \hat{n}_i' \times (\nabla' \times \mathbf{E}_0(\mathbf{s}_\beta)) \right] \end{array} \right\}, \tag{2.18}$$

where $\mathbf{C}_i^{\mathbf{s}_\beta} = \{(C_i^{\mathbf{s}_\beta})_n^m\}$ with $i=0,1,...,J$ and $n=0,1,2,...p-1;\ m=-n,-n+1,...,n-1,n.$ As each box may contain more than one elements, their coefficients are grouped together with respect to different poles as

$$(C_i^{(l_{\max},q)})_n^m = \sum_{\mathbf{s}_\beta \in (l_{\max},q)} (C_i^{\mathbf{s}_\beta})_n^m. \tag{2.19}$$

Then $\mathbf{C}_i^{(l_{\max},q)} = \{(C_i^{(l_{\max},q)})_n^m\}$ is the coefficient vector of box $(l_{\max},q)$. We notice that the expression of the interaction (Eq. (2.18)) involves only the information of the source element $\mathbf{s}_{\beta}$ .

*Step 2. Upward pass*

After collecting the initial interactions at the finest level, they are then translated from box $(l,q)$ to its parent box $(l-1,q')$ (see the Appendix C) by applying a singular-base-to-singular-base translation which we denote by S|S [Gumerov and Duraiswami, 2004]. This means we re-expand the Green's function at the parent box's center $\mathbf{r}^{\star}_{(l-1,q')}$. The translated coefficient vectors are then summed in the parent boxes as

$$\mathbf{C}_i^{(l-1,q')} = \sum_{(l,q) \in \text{Children}(l-1,q')} (\mathbf{S}_i \,|\, \mathbf{S}_i)(\mathbf{t}) \cdot \mathbf{C}_i^{(l,q)},\ i=0,1,...,J. \tag{2.20}$$

Here $\mathbf{t} = \mathbf{r}^{\star}_{(l-1,q')} - \mathbf{r}^{\star}_{(l,q)}$ is the translation vector and $(\mathbf{S}|\mathbf{S})(\mathbf{t})$ is the translation matrix that operates on coefficient vector $\mathbf{C}_i^{(l,q)}$ and produces $\mathbf{C}_i^{(l-1,q')}$. "Children$(l-1,q')$" denotes the children boxes of box $(l-1,q')$ (see the Appendix C). The upward pass is repeated hierarchically from level $l=l_{\max}$ to $l=2$. The interactions from the elements are finally passed to and grouped in a few large boxes on $l=2$.

*Step 3. Downward pass*

In the downward pass, two kinds of translations are conducted. First, the interactions of the source boxes are translated to the receiver boxes on the same level by applying S|R (singular-base-to-regular-base) translations [Gumerov and Duraiswami, 2004]. This is done by transforming the far-field expansions into the near-field expansions. Since the far-field expansion is not applicable to all the source boxes (as indicated in Eq. (2.17)), the S|R translation is only applied to the far-field FF($l$,$q$) boxes (see the Appendix C for details). Second, we translate the interactions from the parent box to its children box through R|R (regular-base-to-regular-base) translations [Gumerov and Duraiswami, 2004]. In the end, we sum the interactions from both the FF($l$,$q$) boxes and the parent boxes as

$$
\begin{aligned}
\mathbf{D}_i^{(l,q)} = & \sum_{(l-1,q')=\mathrm{Parent}(l,q)} (\mathbf{R}_i \,|\, \mathbf{R}_i)(\mathbf{t}) \cdot \mathbf{C}_i^{(l-1,q')} \\
& + \sum_{(l,q'')\in \mathrm{FF}(l,q)} (\mathbf{S}_i \,|\, \mathbf{R}_i)(\mathbf{t}) \cdot \mathbf{C}_i^{(l,q'')}, \; i = 0,1,...,J,
\end{aligned}
\tag{2.21}
$$

where Parent($l$,$q$) denotes the parent box of ($l$,$q$) (see the Appendix C). The above procedure is repeated from level $l = 2$ to the finest level.

*Step 4. Distribution and final summation*

After the downward-pass step, the interactions have been passed to each box on the finest level. In the final step, these interactions are distributed to each element contained in the box, followed by a summation of the near-field interactions. The near-field interactions are the direct interactions from the elements of the neighbor boxes (see the Appendix C) and from the elements contained in the same box (including itself). We denote the near-field elements with respect to the element $\mathbf{s}_\alpha$ ( $\mathbf{s}_\alpha \in (l_{\max}, q)$ ) as $\mathrm{NF}(\mathbf{s}_\alpha) \in [\mathrm{Neighbor}(l_{\max}, q) \bigcup (l_{\max}, q)]$ . The final result of the matrix-vector multiplication on the left-hand side (LHS) of Eq. (2.9) is obtained as:

$$
\begin{bmatrix} \bar{\mathbf{\Gamma}} & \bar{\mathbf{\Lambda}} \end{bmatrix} \begin{bmatrix} \mathbf{\Phi} \\ \mathbf{\Psi} \end{bmatrix} = \begin{bmatrix} \mathbf{\Phi}_0^{LHS} \\ \mathbf{\Phi}_1^{LHS} \\ \vdots \\ \mathbf{\Phi}_J^{LHS} \end{bmatrix}, \; \mathbf{\Phi}_i^{LHS} = \begin{bmatrix} \mathbf{\Phi}_{i,1}^{LHS} \\ \mathbf{\Phi}_{i,2}^{LHS} \\ \vdots \\ \mathbf{\Phi}_{i,u_i}^{LHS} \end{bmatrix},
\tag{2.22}
$$

where

$$
\begin{aligned}
\mathbf{\Phi}_{i,\alpha}^{LHS} \Big|_{\mathbf{s}_\alpha \in (l_{\max},q)} = & \, \mathbf{D}_i^{(l_{\max},q)} \cdot \mathbf{R}_i(\mathbf{s}_\alpha - \mathbf{r}^{\star}_{(l_{\max},q)}) \\
& + \sum_{\mathbf{s}_\beta \in \mathrm{NF}(\mathbf{s}_\alpha)} \left\{ \begin{matrix} \bar{\mathbf{\Gamma}}_{i,\alpha\beta} \cdot \mathbf{E}_0(\mathbf{s}_\beta) \\ + \bar{\mathbf{\Lambda}}_{i,\alpha\beta} \cdot \left[ \hat{n}_i' \times (\nabla' \times \mathbf{E}_0(\mathbf{s}_\beta)) \right] \end{matrix} \right\}, \\
i = 0,1,2,...,J. &
\end{aligned}
\tag{2.23}
$$

Note here the near-field interactions (the second term on the right-hand side of Eq. (2.23)) are calculated by using Green's function directly and $\overline{\boldsymbol{\Gamma}}_{i,\alpha\beta}$ and $\overline{\boldsymbol{\Lambda}}_{i,\alpha\beta}$ are given by Eq. (2.12).

### 2.3. *Boundary integral equations for good conductors*

For scattering system consisting of good conductors, the fields in the scatters undergo dramatic oscillations within a layer of skin depth. Hence, the direct modeling with the above method would become impractical due to the need of a huge number of discretizing elements. However, we note the fields in the open domain can vary less rapidly. In this case, the impedance boundary condition (IBC) [Chew *et al.*, 2001; Tsang *et al.*, 2001] can be utilized in the FMBEM, which states

$$\hat{n}\times\mathbf{E} = Z\hat{n}\times(\hat{n}\times\mathbf{H}). \tag{2.24}$$

It relates the tangential components of the electric field and the magnetic field through the surface impedance $Z=\sqrt{\mu/\varepsilon}$ of the media. By applying this boundary condition, one can reformulate Eq. (2.5) with only the normal and tangential components of the electric field on the boundary. The sacrifice is that we have to omit the modeling of the inside domains, which is not that important in many applications.

Inserting Eq. (2.24) into Eq. (2.5), we obtain the electric field integral equations with IBC as

$$\begin{aligned}
&\oint_{\partial\Omega_0}^{\mathrm{CPV}} \mathrm{d}A' \left\{ \begin{aligned} &\left[\hat{n}_0'\nabla' g_0(\mathbf{s}',\mathbf{s}) - \hat{n}_0'\cdot\nabla' g_0(\mathbf{s}',\mathbf{s})\overline{\mathbf{I}} - \nabla' g_0(\mathbf{s}',\mathbf{s})\hat{n}_0'\right]\cdot\mathbf{E}_0(\mathbf{s}') \\ &-\frac{i\omega\mu_0}{Z} g_0(\mathbf{s}',\mathbf{s})\overline{\mathbf{I}}\cdot\left[\hat{n}_0'\times(\hat{n}_0'\times\mathbf{E}_0(\mathbf{s}'))\right] \end{aligned} \right\} \\
&= \frac{1}{2}\mathbf{E}_0(\mathbf{s}) - \mathbf{E}_0^{\mathrm{inc}}(\mathbf{s}),
\end{aligned} \tag{2.25}$$

The above equation is then numerically implemented in a similar way as in Section 2.2.

## 3. Pre-conditioning, numerical benchmarks and complexity study

### 3.1. *Block-diagonal pre-conditioner*

The discretization of the integral equations in the FMBEM produces a dense asymmetric matrix in Eq. (2.9). Solving such equations by an iterative method like fGMRES is usually very challenging due to its slow convergence. A remedy is to employ a pre-conditioner that is sparse enough to be applied efficiently and also partially resembles the original matrix. A right pre-conditioner $\mathbf{M}$ is defined as

$$\mathbf{A}\mathbf{M}^{-1}\mathbf{M}\mathbf{x} = \mathbf{b}, \tag{3.1}$$

where instead of solving the equation $\mathbf{A}\mathbf{x}=\mathbf{b}$, we solve the right-preconditioned system $\overline{\mathbf{A}}\overline{\mathbf{x}}=\mathbf{b}$, where $\overline{\mathbf{A}}=\mathbf{A}\mathbf{M}^{-1}$ and $\overline{\mathbf{x}}=\mathbf{M}\mathbf{x}$. The new system should have a better eigen spectrum so that a better convergence is expected. In the FMBEM, matrix $\mathbf{A}$ can be decomposed as

$$\mathbf{A} = \mathbf{A}_{\text{self}} + \mathbf{A}_{\text{near}} + \mathbf{A}_{\text{far}}, \tag{3.2}$$

where $\mathbf{A}_{\text{self}}$ represents the self-interactions that account for the diagonal elements in each block; $\mathbf{A}_{\text{near}}$ is the interactions from the nearby elements contained in the same box and in the neighbor boxes; $\mathbf{A}_{\text{far}}$ represents the interactions originating from the far-field boxes. We construct the block-diagonal pre-conditioner as $\mathbf{M} = \mathbf{A}_{\text{self}}$, which only takes account of the diagonal elements of the self-interactions.

The whole iterative procedure is divided into two parts: an outer main fGMRES loop and an inner GMRES [Saad and Schultz, 1986] loop. The inner loop with matrix $\mathbf{M}$ helps to precondition the matrix. We set an upper limit of 10 for the total number of the inner iterations. To guarantee a cheap application of the pre-conditioning, we also employ a relatively larger convergence tolerance for the inner loop. Figure 3 shows the comparison of the convergence speeds between the FMBEM with and without the right pre-conditioner, where we calculated a plane wave scattered by a dielectric sphere ( $\varepsilon_r = 1.33^2$ ) of radius $a$ = 1.0 with $ka$ = 1.67 and $ka$ = 5.57 ($k$ is the wavenumber in the sphere). In both cases, the pre-conditioning substantially reduces the total number of iterations.

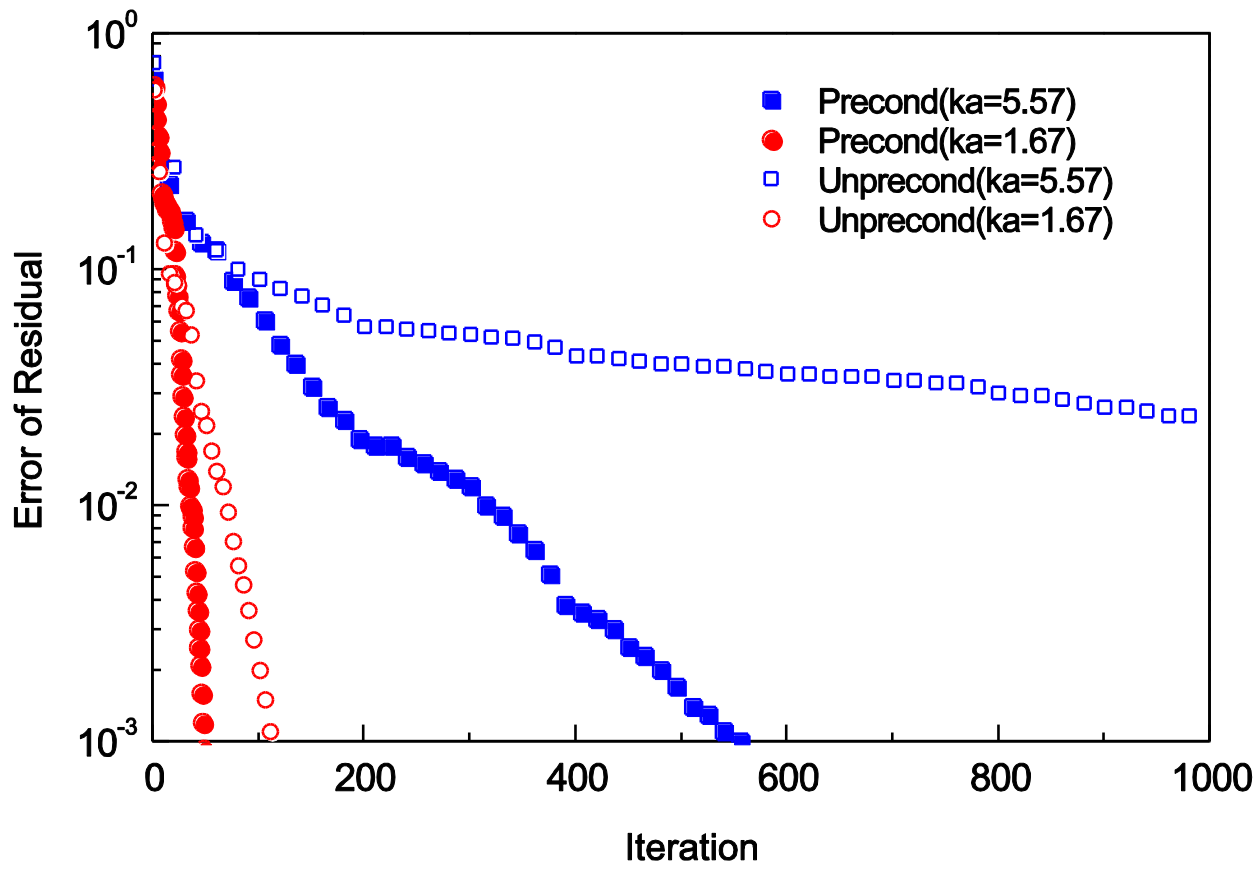

Fig. 3. Convergence speed comparisons between the FMBEM with and without the right-pre-conditioner. The cases here correspond to a plane wave scattered by a dielectric sphere of $ka = 1.67$ and $ka = 5.57$ . A mesh of $N$=7040 is used in the former case while $N$=28160 is used in the latter case. In the second case, the un-preconditioned one does not converge.

### 3.2. *Numerical tests and comparisons with the Mie theory*

To benchmark the accuracy of the FMBEM we calculate the scattering of a plane wave by dielectric/metallic spheres and compare the results with Mie scattering theory results. The dielectric spheres have fixed radii of $a = 1.0$, relative permittivity $\varepsilon_r = 1.33^2$ and relative permeability $\mu_r = 1.0$. The incident plane wave is defined as $\mathbf{E}_{\text{inc}} = \hat{x} \cdot E_0 e^{i(k_0 z - \omega t)}$.

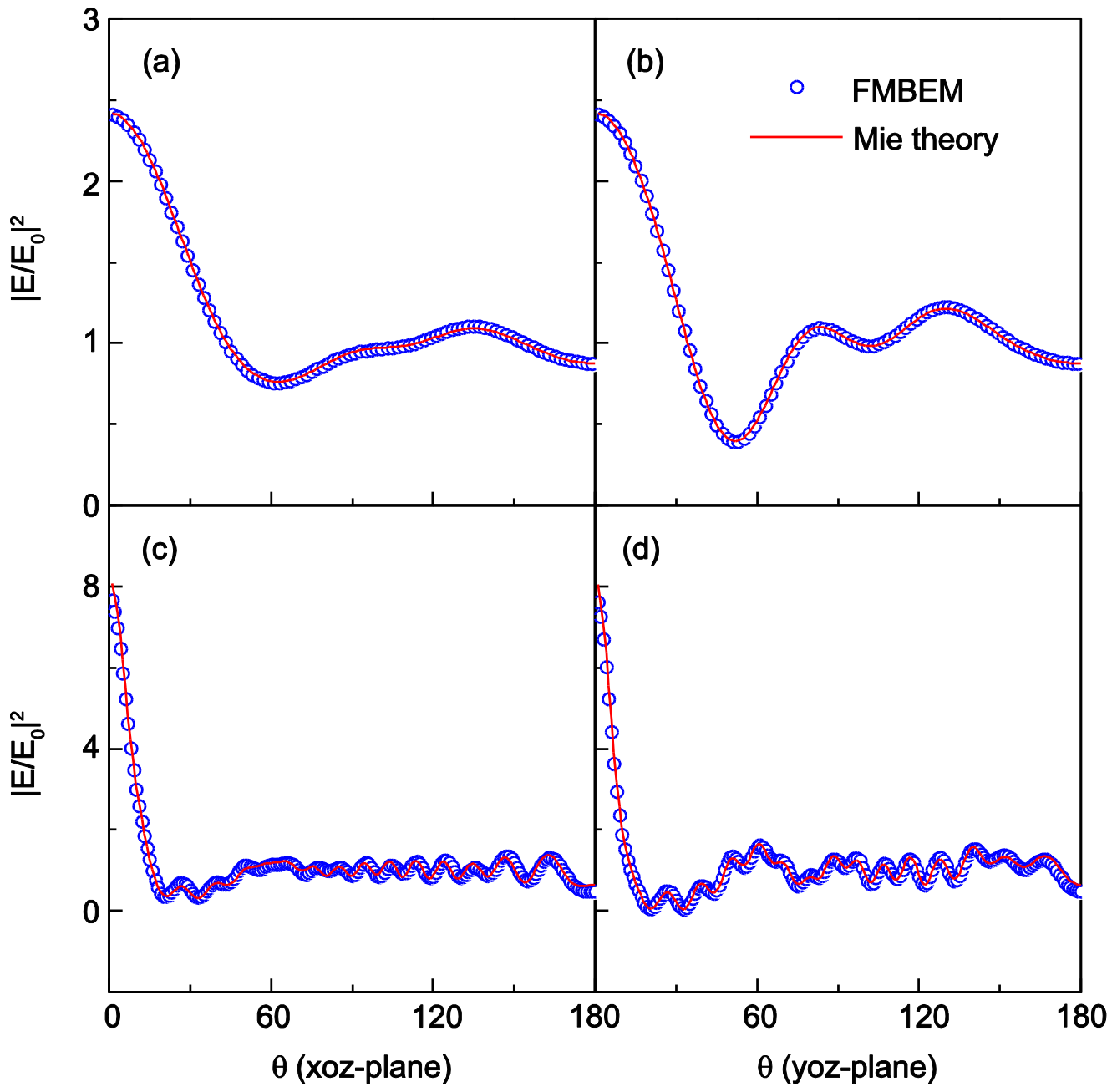

Fig. 4. Comparison of the near field between the FMBEM and Mie theory for a dielectric sphere of $ka = 3.34$ ((a),(b)) and $ka = 16.71$ ((c),(d)) under the plane wave incidence. A mesh of $N$=3374 is used for the former case while a mesh of N=112640 is used for the latter one.

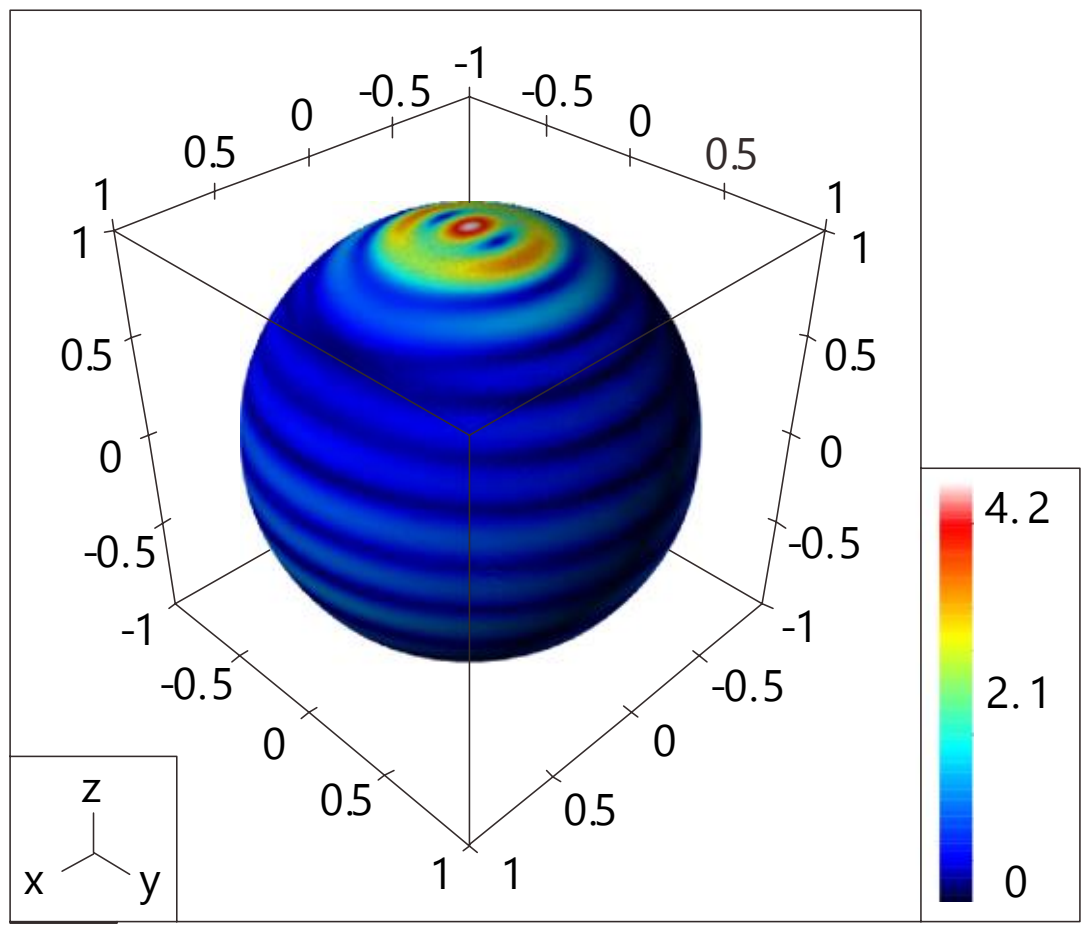

Fig. 5. Electric field magnitude pattern at the surface of the sphere calculated by the FMBEM for $ka = 16.71$. The incident direction of the plane wave is along $z$-axis.

We consider different wavelengths in different cases. In order to test the near-field accuracy of the FMBEM, we consider two cases: a low frequency case with $ka = 3.34$ ($k$ is the wavenumber in the sphere) and a high frequency case with $ka = 16.71$. The results are shown in Fig. 4, where we compare the near-field intensity sampled on a circle of radius $R=2a$ with respect to the sphere's center on both the *xoz*-plane and the *yoz*-plane. The agreement with analytic Mie scattering results is excellent. Figure 5 shows the electric field magnitude on the surface of the dielectric sphere in the case $ka = 16.71$.

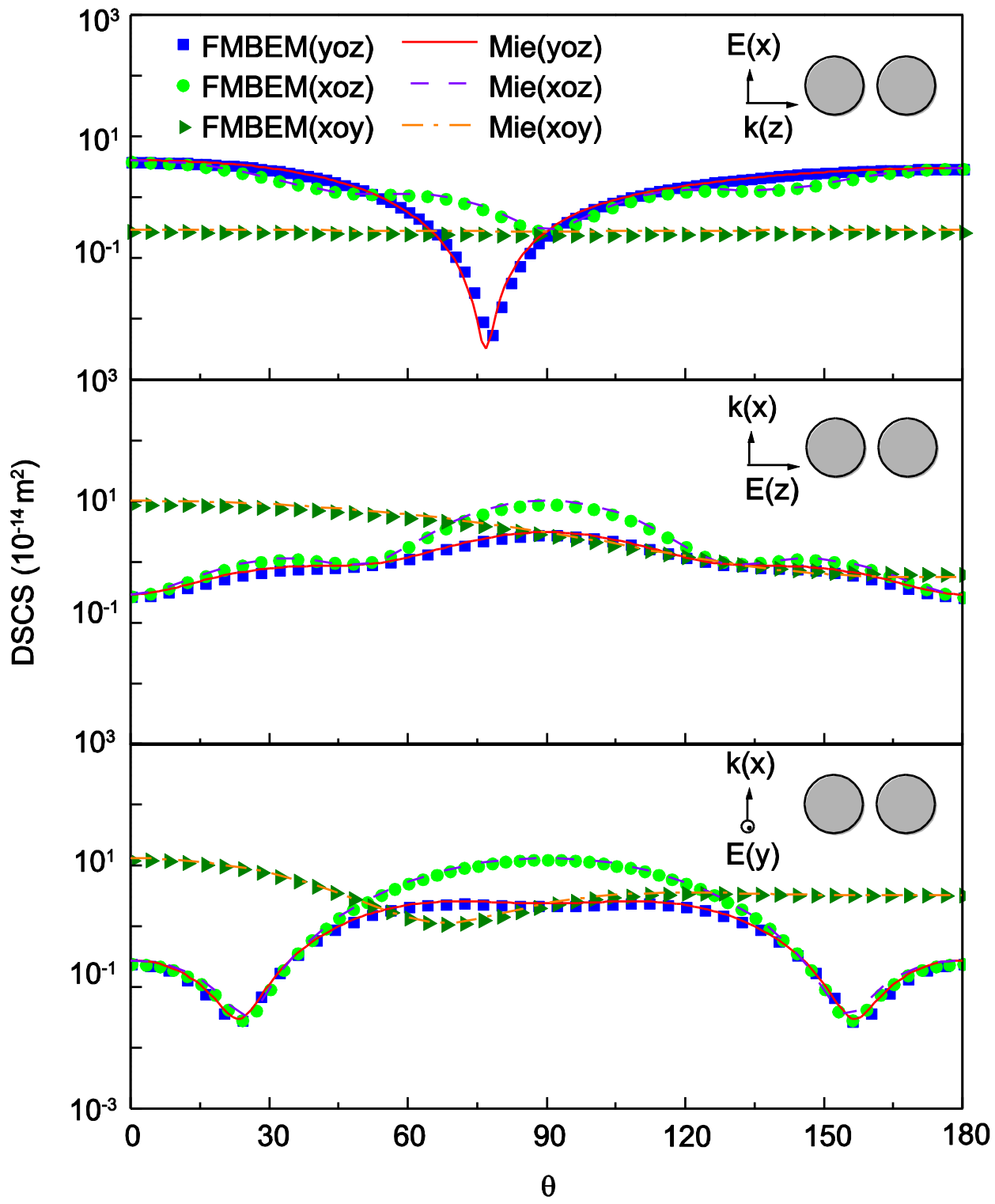

Fig. 6. Comparisons of the DSCS calculated by Mie theory (lines) and by the FMBEM (symbols). The cases studied here are two neighboring gold spheres under the incidence of a plane wave with different configurations (shown by the insets). The two spheres have a radius of 100nm and are separated by a distance of 250 nm. The permittivity of gold is described by the Drude model mentioned in Section 3.2 and the incident wavelength is 500 nm. A mesh of $N$=11904 is used in all the three cases.

To test the accuracy of the FMBEM at the far-field, we computed the differential scattering cross section (DSCS) of two neighboring gold spheres under plane wave incidence. The relative permittivity of gold is described by the Drude model: $\varepsilon_r = 9.6 - \omega_p^2 / (\omega^2 + i\omega_t \omega)$, where $\omega_p = 1.37\times 10^{16}$ rad/s and $\omega_t = 1.068\times 10^{14}$ rad/s. The two gold spheres have the same radius of 100 nm and are separated by a distance of 250 nm. The incident plane wave has a wavelength of 500 nm. Figure 6 shows the comparisons of the computed DSCS between Mie theory and the FMBEM, where a good match is again evident.

The FMBEM can also handle problems involving good conductors. We consider the so-called "NASA almond" as an example since it is a frequently-used benchmark target for computing radar cross section (RCS) [Dominek *et al.*, 1986; Dominek and Shamanski, 1990; Woo *et al.*, 1993]. The exact mathematical definition of the almond can be found in Woo *et al*. [1993]. The model is about 25.2 cm long, 11.7 cm wide and 3.9 cm high. The computed monostatic RCS at the frequency of 7GHz for the vertical-vertical polarization is shown in Fig. 7. We have taken the relative permittivity of the material to be $1+10^8 i$ so that it essentially represents a very good metal. And the IBC is applied in this case. The red-solid line is the FMBEM result, which shows a good match with the experimental result represented by the blue-dashed line.

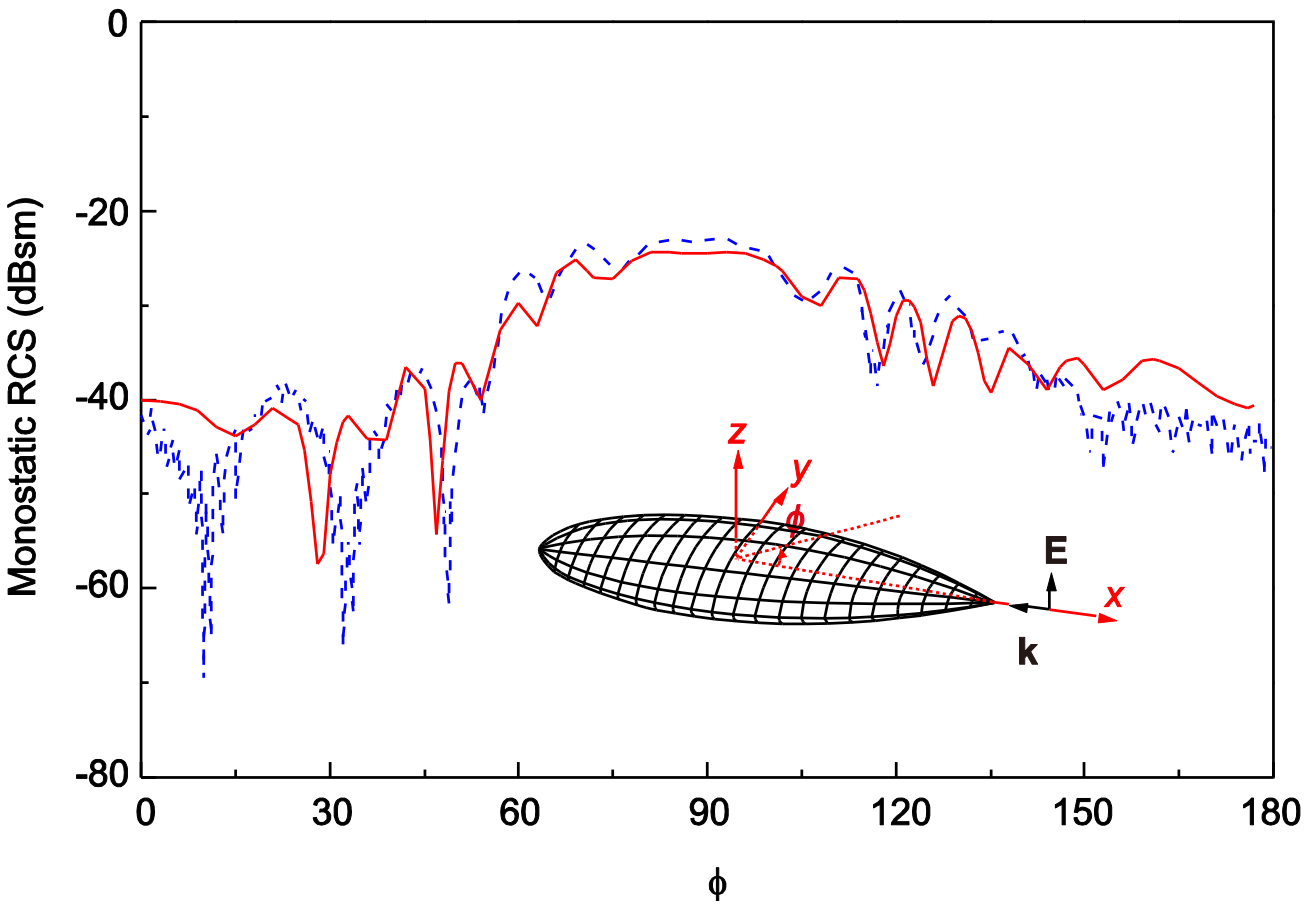

Fig. 7. Monostatic RCS of the "NASA almond" at 7 GHz for the vertical-vertical polarization. The red-solid line is the result given by the FMBEM while the blue-dotted line is the experimental results from Dominek *et al*. [1990].

### 3.3. *Computational complexity*

The complexity of the FMBEM is related to the total number of elements *N*, which is determined by a dimensionless parameter *kD*. Here *k* is the wave number in the media and *D* is the computational domain size. If the size of an element is approximately *d*, then $N \sim D^2/d^2$. To guarantee an accurate solution the constraint $d/\lambda \ll 1$ must be applied. Let us define a parameter $\eta = d/\lambda$ so that $d = 2\pi\eta/k$. For the case of a sphere, we have $N = \pi D^2/d^2 = (kD)^2/(4\pi\eta^2)$ .To obtain accurate results, we set $\eta = 0.05$, which guarantees about 20 elements per wavelength. Figure 8 shows the complexities of the current algorithm, where we considered a plane wave scattered by a dielectric sphere of radius *a* = 1.0 and $\varepsilon_r = 1.33^2$. All the computations are done on a single thread of an Intel Xeon E5640 (2.67GHz) CPU. The fitting shows the time per iteration approximately scales as $O[(kD)^2 \log(kD)^2] = O(N\log N)$ , which means the total solution time of the FMBEM is $O(n_{\text{iter}} N \log N)$ with $n_{\text{iter}}$ being the total number of iterations (we used a constant mesh for the range of $kD \in [10^{-1}, 10^1]$ ). The memory complexity of the FMBEM scales as $O(kD)^2 = O(N)$ theoretically. However, due to various possibilities of memory management in the implementation, the actual memory scaling may deviate a little bit from the ideal linear law. In the frequency range we studied, our algorithm scales approximately as $O(kD)^{2.48} = O(N^{1.24})$. This makes it possible to solve very large systems involve millions of unknowns on a single desktop PC.

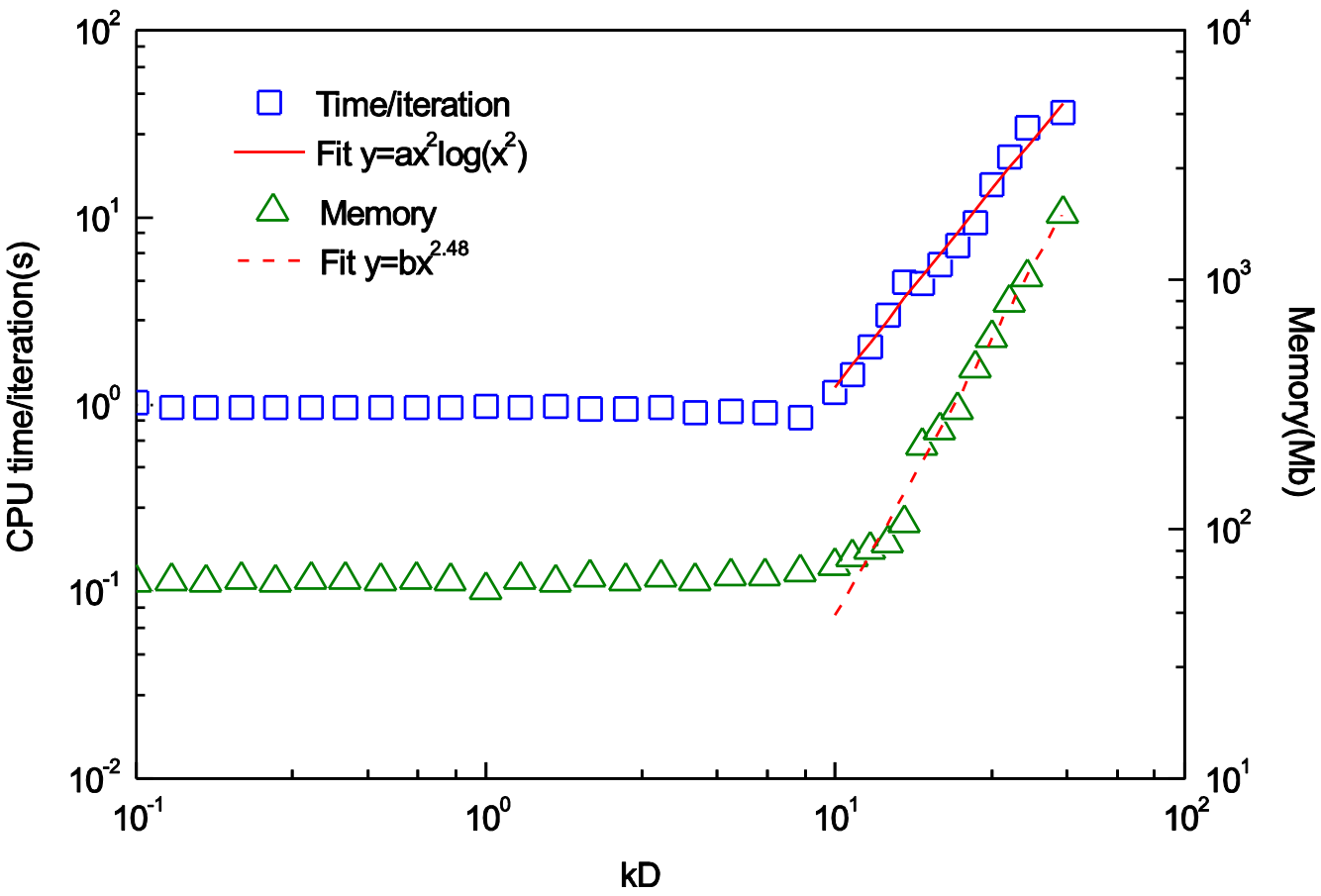

Fig. 8. Complexities of the CPU time per iteration (squares) and the memory consumed (triangles) of the FMBEM for solving a dielectric sphere scattering problem. We use a constant number of meshes for $kD \le 10$.

## 4. Applications in the study of plasmonic systems and optical forces

### 4.1. *Plasmonic properties of a torus and a split-ring resonator*

We apply the developed algorithm to investigate the properties of a plasmonic torus structure (Fig. 9 and Fig. 10) and a SRR (Fig. 11 and Fig. 12). Both of the two structures have an inner radius of 20 nm and an outer radius of 50 nm. The SRR structure has an air gap of $\pi/10$ along $x$-axis. The material of the structures is taken to be silver whose permittivity in the visible range is described by a Drude model plus two Lorentz pole functions [McMahon *et al.*, 2009]:

$$\varepsilon_r = \varepsilon_\infty - \frac{\omega_D^2}{\omega^2 + i\omega\gamma_D} + \sum_{L=1}^{2} \frac{\Delta\varepsilon_L \omega_L^2}{\omega_L^2 - 2i\omega\delta_L - \omega^2}, \tag{4.1}$$

where the parameters are fitted to experimental data [Johnson and Christy, 1972] and take the values: $\varepsilon_\infty = 1.17152$, $\omega_D = 1.39604\times10^{16}$, $\gamma_D = 12.6126$, $\Delta\varepsilon_1 = 2.23994$, $\Delta\varepsilon_2 = 0.222651$, $\omega_1 = 8.25718\times10^5$, $\omega_2 = 3.05707\times10^{15}$, $\delta_1 = 1.95614\times10^{14}$ and $\delta_2 = 8.52675\times10^{14}$. Here a unit of “rad/s” is assumed for $\omega_D, \gamma_D, \omega_L$ and $\delta_L$. The absorption cross section, scattering cross section and total extinction cross section of the two structures can be calculated as [Jackson, 1999],

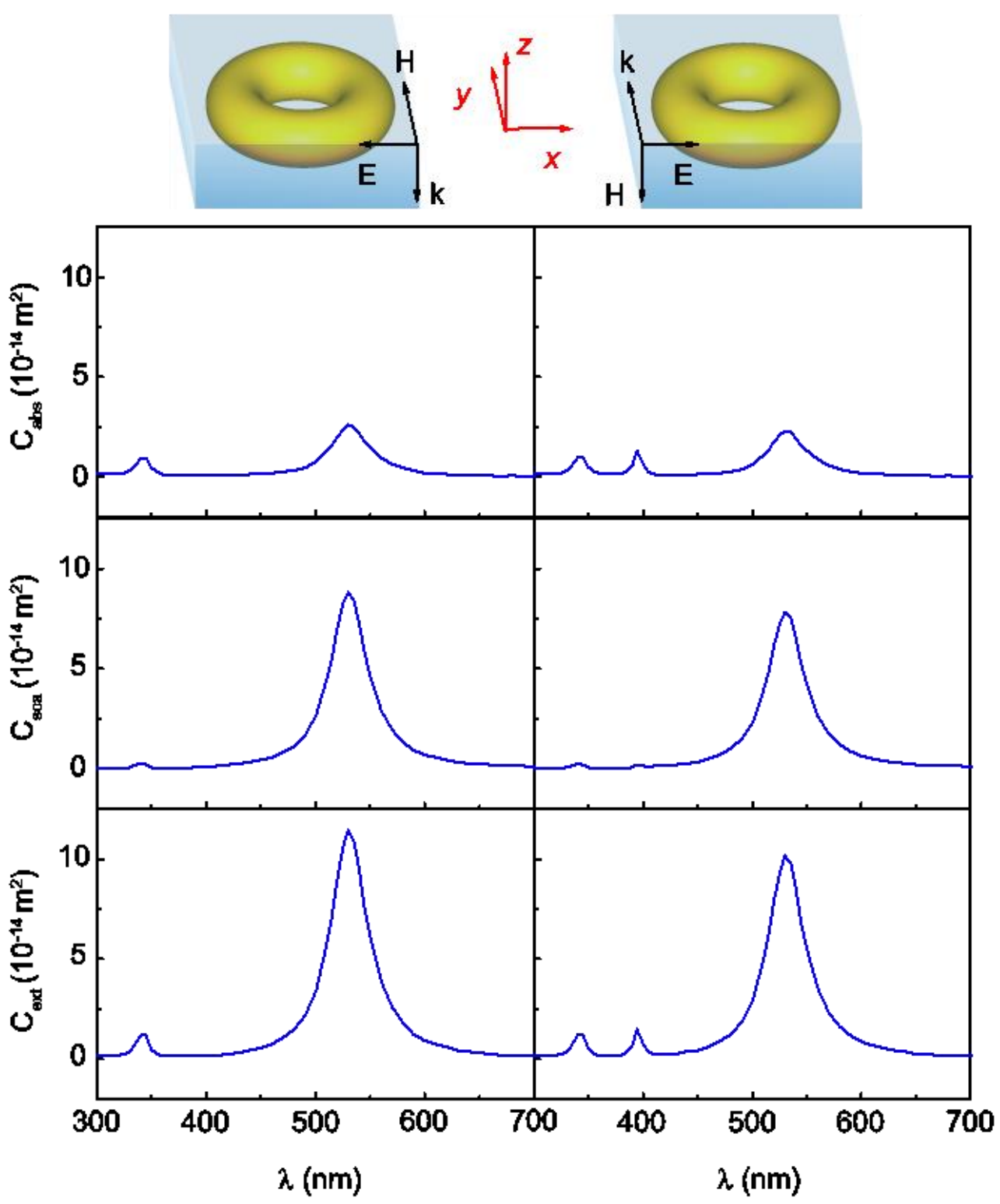

Fig. 9. Absorption cross section ($C_{abs}$), scattering cross section ($C_{sca}$) and extinction cross section ($C_{ext}$) of a silver torus under different polarizations of the incident plane wave. The torus has an inner radius of $r = 20$ nm and an outer radius of $R = 50$ nm .

$$\begin{aligned}
C_{\text{abs}} &= \frac{P_{\text{abs}}}{I_{\text{inc}}} = -\frac{Z_0}{\left|E_0\right|^2}\oint_S \text{Re}(\mathbf{E}\times\mathbf{H}^*)\cdot\hat{n}\text{d}A, \\
C_{\text{sca}} &= \frac{P_{\text{sca}}}{I_{\text{inc}}} = \frac{Z_0}{\left|E_0\right|^2}\oint_S \text{Re}(\mathbf{E}_{\text{sca}}\times\mathbf{H}^*_{\text{sca}})\cdot\hat{n}\text{d}A, \qquad (4.2)\\
C_{\text{ext}} &= C_{\text{abs}} + C_{\text{sca}}.
\end{aligned}$$

Here $P_{\text{abs}}$ and $P_{\text{sca}}$ are the absorbed power and the scattered power, respectively. $I_{\text{inc}}$ is the incident power per unit area. $Z_0 = \sqrt{\mu_0/\varepsilon_0}$ is the impedance of vacuum. $E_0$ is the amplitude of the incident electric field. **E** and **H** are the total electric field and magnetic field while $\mathbf{E}_{\text{sca}}$ and $\mathbf{H}_{\text{sca}}$ are the scattered electric field and magnetic field, respectively. The integrals are done over a closed surface surrounding the scatter.

Figure 9 shows the calculated results of the torus structure by the FMBEM, where two different polarizations of the incident plane wave are considered. For the $E_x$-$H_y$ polarization, two resonances emerge at $\lambda = 342$ nm and $\lambda = 530$ nm . For the $E_x$-$H_z$ polarization, a third resonance can be found at $\lambda = 395$ nm in addition to the two resonances of the former case. To reveal the physics here, we plot in Fig. 10 the electric field patterns corresponding to the three resonances in the $E_x$-$H_z$ cases, where both the near-field pattern and far-field pattern are shown. We note the resonances here may involve electric resonances and magnetic resonances at the same time, which make the situation much more complicated. To retrieve this information, we expand the electric scattered field at the far-field zone on the basis of the vector spherical harmonics $\mathbf{N}_n^m$ and $\mathbf{M}_n^m$, which in spherical coordinates are defined as [Bohren and Huffman, 1983]

$$\begin{aligned}\mathbf{N}_n^m &= \frac{h_n(\rho)}{\rho}\cos(m\phi)n(n+1)P_n^m(\cos\theta)\hat{e}_r \\ &\quad + \cos(m\phi)\frac{dP_n^m(\cos\theta)}{d\theta}\frac{1}{\rho}\frac{d\left[\rho h_n(\rho)\right]}{d\rho}\hat{e}_\theta \\ &\quad - m\sin(m\phi)\frac{P_n^m(\cos\theta)}{\sin\theta}\frac{1}{\rho}\frac{d\left[\rho h_n(\rho)\right]}{d\rho}\hat{e}_\phi, \\ \mathbf{M}_n^m &= \frac{m}{\sin\theta}\cos(m\phi)P_n^m(\cos\theta)h_n(\rho)\hat{e}_\theta \\ &\quad - \sin(m\phi)\frac{dP_n^m(\cos\theta)}{d\theta}h_n(\rho)\hat{e}_\phi.\end{aligned} \tag{4.3}$$

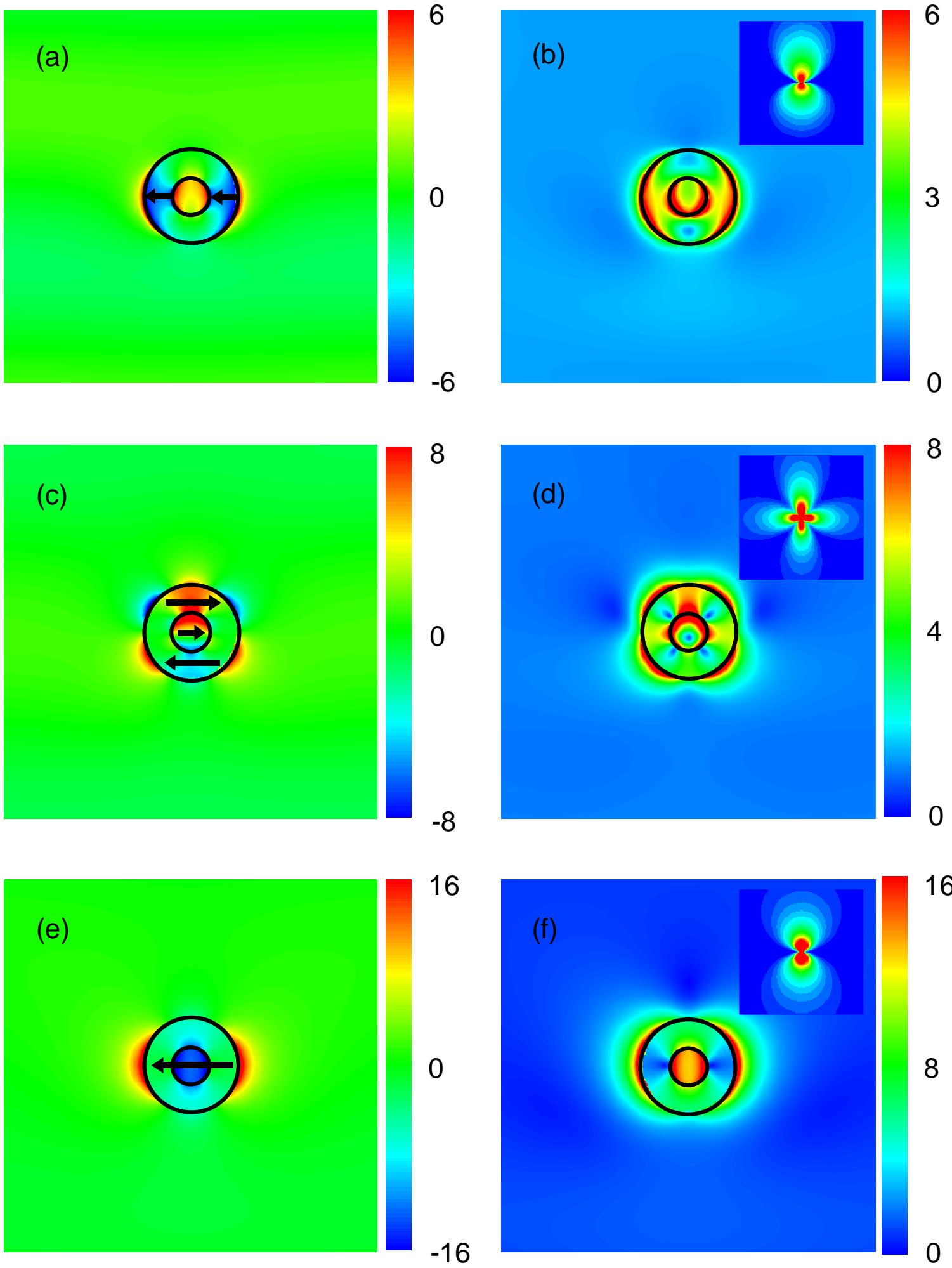

Fig. 10. Electric field patterns of the torus structure at the wavelengths $\lambda = 342$ nm, 395 nm and 530 nm , where (a), (c) and (e) show the *x*-component of the total electric field on *xoy*-plane while (b), (d) and (f) show the amplitude of the total electric field at *xoy*-plane. The insets in panels (b), (d), (f) show the amplitude patterns of the scattered electric field at the far-field. The arrows in panels (a), (c), (e) indicate the direction of the excited electric dipoles.

Table 1. Far-field modes expansion coefficients of the torus structure.

| n | λ = 342 nm | | λ = 395 nm | | λ = 530 nm | |
|---|---|---|---|---|---|---|
| | $\|a_n\|$ | $\|b_n\|$ | $\|a_n\|$ | $\|b_n\|$ | $\|a_n\|$ | $\|b_n\|$ |
| 1 | 0.997 | 0.049 | 0.807 | 0.104 | 1.000 | 0.008 |
| 2 | 0.059 | 0.005 | 0.581 | 0.018 | 0.001 | 0.009 |
| 3 | 0.002 | 0.000 | 0.001 | 0.004 | 0.000 | 0.000 |

Here $\mathbf{N}_n^m$ and $\mathbf{M}_n^m$ correspond to the electric type and magnetic type normal modes, respectively. $P_n^m$ is the associated Legendre function and $h_n$ is the spherical Hankel function of the first kind. Under the plane wave incidence, we have $m = 1$. The electric scattered field can be expressed as

$$\mathbf{E}_{\text{sca}} = \sum_{n=1}^{\infty}\left(a_n\mathbf{N}_n^m + b_n\mathbf{M}_n^m\right), \tag{4.4}$$

where the expansion coefficients can be calculated as

$$a_n = \frac{\int_0^{2\pi}\int_0^{\pi}\mathbf{E}_{\text{sca}}\cdot\mathbf{N}_n^m\sin\theta\,\mathrm{d}\theta\,\mathrm{d}\phi}{\int_0^{2\pi}\int_0^{\pi}\left|\mathbf{N}_n^m\right|^2\sin\theta\,\mathrm{d}\theta\,\mathrm{d}\phi}, b_n = \frac{\int_0^{2\pi}\int_0^{\pi}\mathbf{E}_{\text{sca}}\cdot\mathbf{M}_n^m\sin\theta\,\mathrm{d}\theta\,\mathrm{d}\phi}{\int_0^{2\pi}\int_0^{\pi}\left|\mathbf{M}_n^m\right|^2\sin\theta\,\mathrm{d}\theta\,\mathrm{d}\phi}. \tag{4.5}$$

The obtained expansion coefficients are shown in Table 1, where we have only considered the lowest three orders and the coefficients are normalized to $\sqrt{\sum_{n=1}^{3}|a_n|^2+|b_n|^2}$. We note the resonances at $\lambda = 342$ nm and $\lambda = 530$ nm are mainly electric dipole resonances as the $n = 1$ mode dominates. However, the 342 nm resonance comprises two small electric dipoles combined in a row while the 530 nm resonance is a larger electric dipole, as shown by the arrows in Fig. 10(a) and Fig. 10(e). The resonance at $\lambda = 395$ nm, as indicated by Table 1, is mainly consist of two types of electric resonances: an electric dipole type and an electric quadrupole type. There is also a small magnetic dipole projection.

Figure 11 shows the absorption cross section, scattering cross section and extinction cross section of the SRR structure. Like the torus case, there are also three resonances emerged in the frequency range considered: a third resonance exists in the $E_y$-$H_z$ polarization beside the two resonances in the $E_y$-$H_x$ case. The electric field patterns corresponding to these resonances are shown in Fig. 12 and the retrieved far-field expansion coefficients are listed in Table 2. We note the peak at $\lambda = 342$ nm is formed by the combination of an electric dipole resonance and a magnetic dipole resonance. The resonances corresponding to $\lambda = 384$ nm is mainly an electric dipole resonance, where the electric quadrupole resonance is undermined by the symmetry-breaking due to the existence of the gap. The resonance at $\lambda = 470$ nm is an electric dipole type.

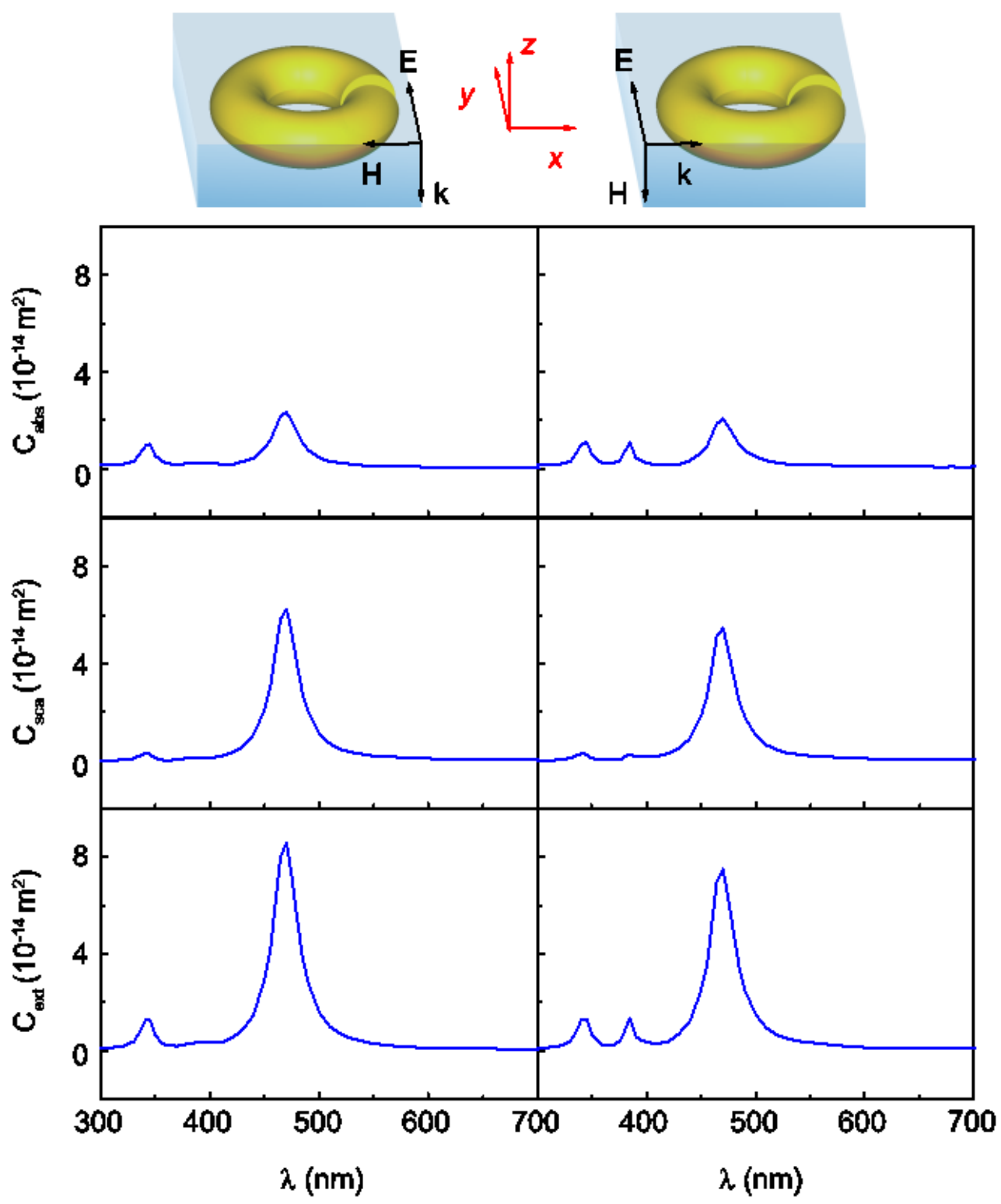

Fig. 11. Absorption cross section ($C_{abs}$), scattering cross section ($C_{sca}$) and extinction cross section ($C_{ext}$) of a SRR under different polarizations of the incident plane wave. The SRR has an inner radius of $r = 20$ nm and an outer radius of $R = 50$ nm . The gap is created by cutting an angle of $\pi / 10$ with respect to the center from the corresponding torus structure.

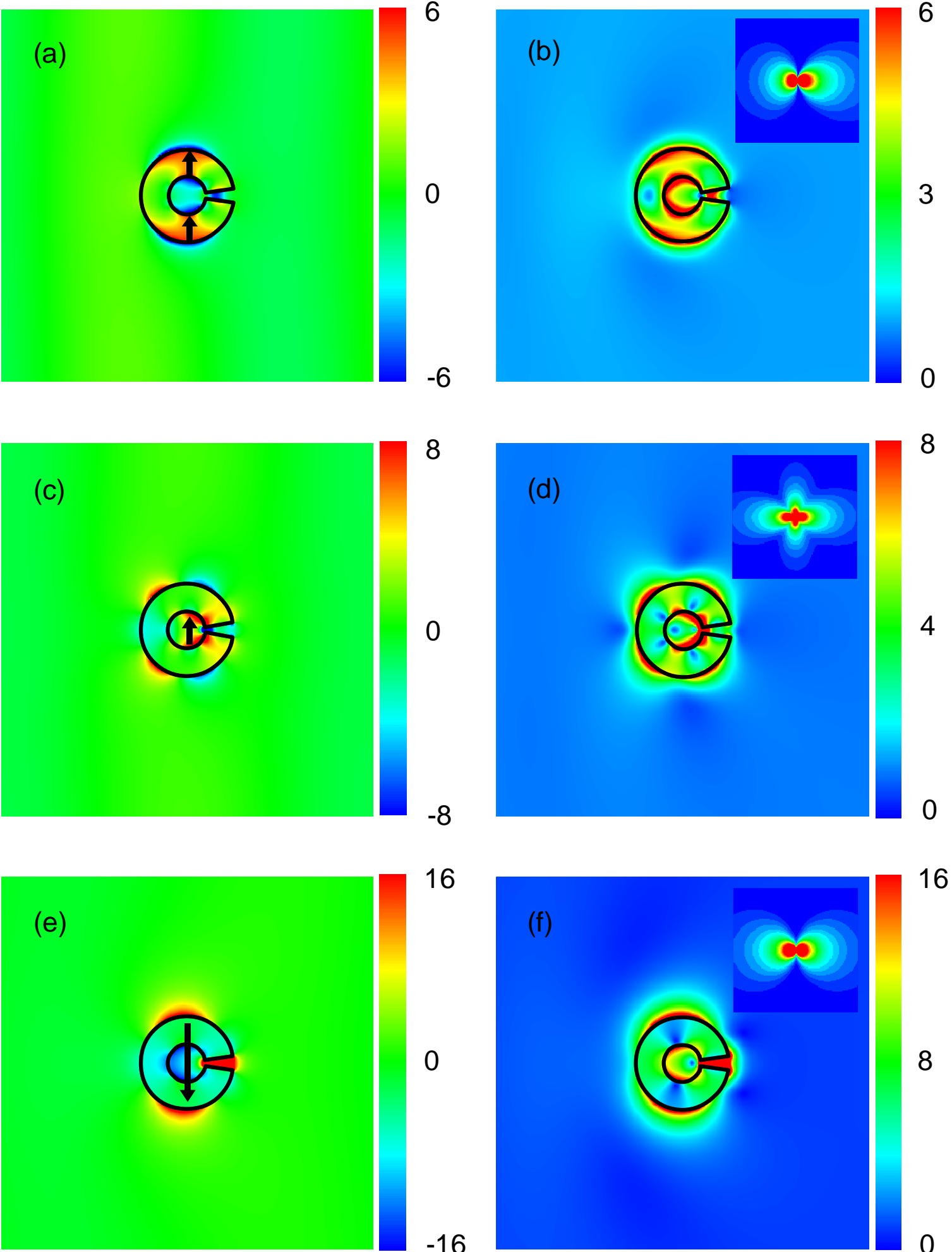

Fig. 12. Electric field patterns of the SRR structure at the wavelengths $\lambda = 342$ nm, 384 nm and 470 nm, where (a), (c) and (e) show the *x*-component of the total electric field on *xoy*-plane while (b), (d) and (f) show the amplitude of the total electric field at *xoy*-plane. The inserted figures show the amplitude patterns of the scattered electric field at the far-field. The arrows in panels (a), (c), (e) indicate the direction of the excited electric dipoles.

Table 2. Far-field modes expansion coefficients of the SRR structure.

| n | $\lambda = 342$ nm | | $\lambda = 384$ nm | | $\lambda = 470$ nm | |
|---|---|---|---|---|---|---|
| | $\lvert a_n \rvert$ | $\lvert b_n \rvert$ | $\lvert a_n \rvert$ | $\lvert b_n \rvert$ | $\lvert a_n \rvert$ | $\lvert b_n \rvert$ |
| 1 | 0.868 | 0.492 | 0.996 | 0.078 | 0.980 | 0.179 |
| 2 | 0.064 | 0.019 | 0.046 | 0.002 | 0.008 | 0.002 |
| 3 | 0.010 | 0.001 | 0.007 | 0.000 | 0.001 | 0.000 |

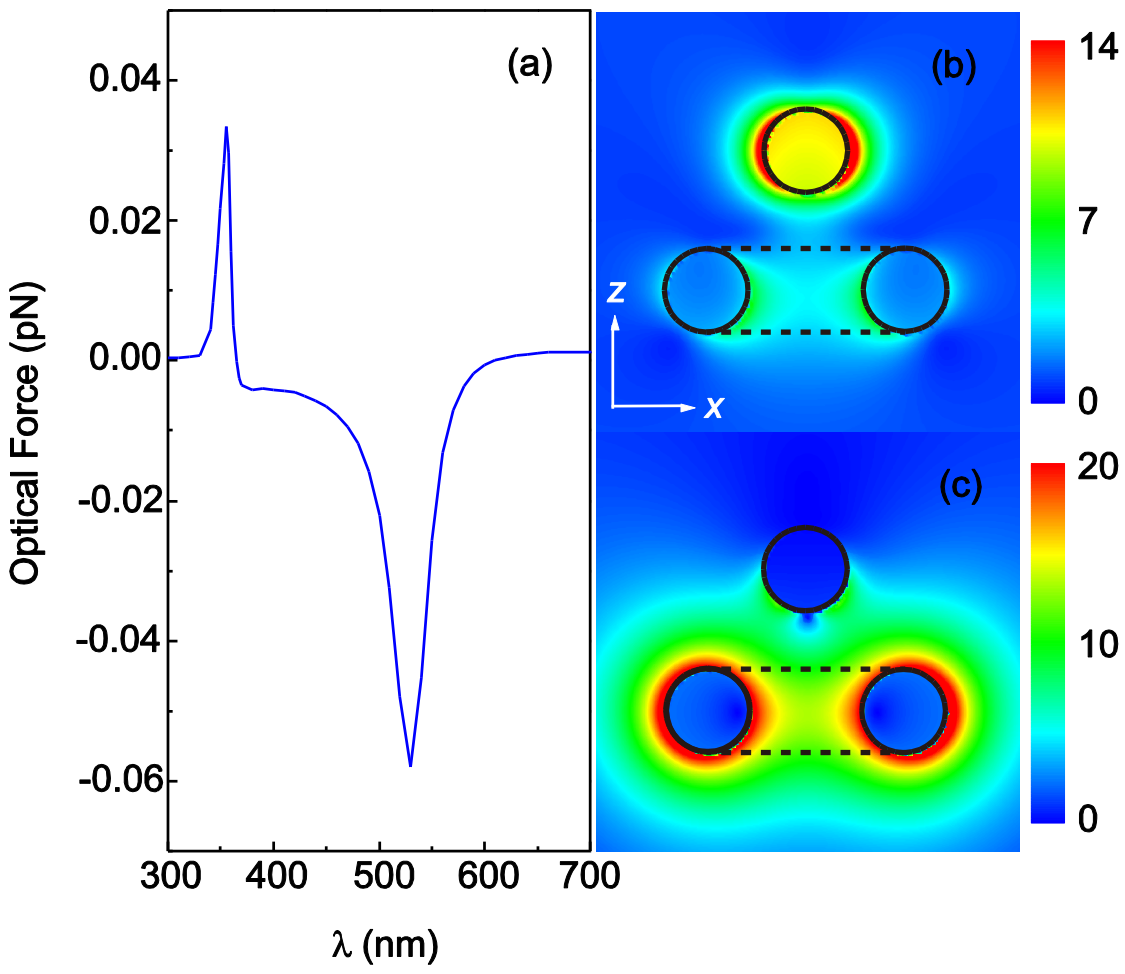

Fig. 13. Optical force between the silver sphere and the silver torus. (a) Optical force along $z$-direction exerted on the silver sphere as a function of the wavelength. (b) Electric field amplitude pattern on the $xoz$-plane at $\lambda = 355$ nm . (c) Electric field amplitude pattern on the $xoz$-plane at $\lambda = 530$ nm .The sphere is 50 nm above the torus' mass center. The permittivity of the silver is described in Section 4.1. The incident plane wave has $k$ along $z$-axis and the polarization along $x$-axis.

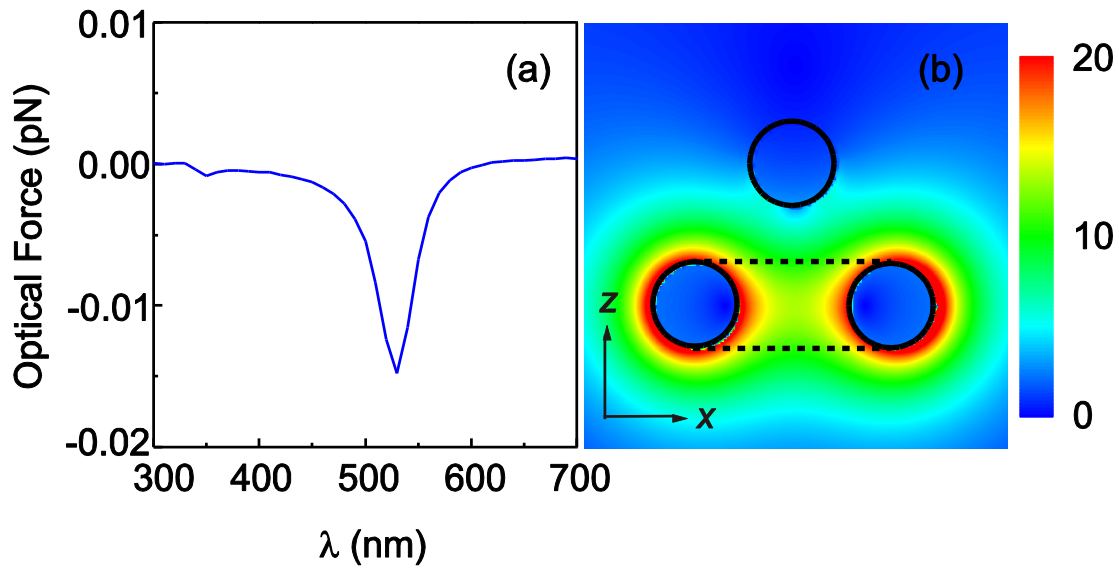

Fig. 14. Optical force between the dielectric sphere ( $\varepsilon_r = 2.4$ ) and the silver torus. (a) Optical force along $z$-direction exerted on the dielectric sphere as a function of the wavelength. (b) Electric field amplitude pattern on the $xoz$-plane at $\lambda = 530$ nm . The sphere is 50 nm above the torus' mass center. The incident plane wave has $k$ along $z$-direction and the polarization along $x$-direction.

### 4.2. *Optical forces and trapping effect of a torus*

An electromagnetic field solver can also be used to compute the light-induced forces using the Maxwell stress tensor approach [Jackson, 1999]. To demonstrate such functionality, we performed a numerical study of optical forces between a silver/dielectric ( $\varepsilon_r = 2.4$ ) sphere and a silver torus excited by an incident plane wave with the intensity of $1\ \text{mW}/\mu\text{m}^2$ . The details of the numerical performance of the FMBEM on calculating optical forces in plasmonic systems are provided in the Appendix D, and the performance is compared to the Mie theory and a commercial finite element method package COMSOL. The radius of the sphere is 15 nm. The torus is the same one as shown in Fig. 9. Figure 13(a) shows the calculated optical force exerted on the silver sphere. The sphere is placed 50 nm above the mass center of the torus and the incident plane wave has *k* along *z*-axis and the polarization along *x*-axis. Two resonance peaks are identified: one at about $\lambda = 355$ nm and the other one at about $\lambda = 530$ nm . Figure 13(b) and Figure 13(c) shows the electric field amplitude patterns on the *xoz*-plane corresponding to the two resonances. There are two kinds of electric dipole resonances: one corresponds to the resonance of the silver sphere, the other one corresponds to the resonance of the silver torus. When the sphere is in resonance, the system can be considered as two parallel electric dipoles (the torus one is very weak) and they repel each other, which accounts for the positive optical force. On the other side, when the torus is in resonance, the sphere is mainly polarized by the strong dipole fields of the torus. So the two dipoles are anti-parallel to each other and attraction happens, which accounts for the "negative" optical force. This shows that a plasmonic torus can either attract or repel a plasmonic particle through light-induced forces. If the plasmonic particle is off-resonance, the dipole in the particle is induced by the near field of the torus and the force is attractive. But if the plasmonic particle is resonating, the induced dipole of the particle and that of torus are in phase and the light-induced force is repulsive. Figure 14(a) shows the calculated optical force corresponding to the dielectric sphere case. Here the configuration is the same except that we replace the silver sphere with a dielectric ( $\varepsilon_r = 2.4$ ) sphere. Since the dipole resonance of the dielectric sphere is far away from the frequency range considered, only the torus dipole resonance is induced by the external field in this case. The dipole moment of the dielectric sphere is induced by the near field of the torus and the force should be attractive. This is indeed the case as shown by the numerically calculated electric field pattern in Fig. 14(b). The dielectric sphere is polarized by the resonance field of the torus dipole and an attractive force is induced.

We also studied the capability of optical trapping in the torus structure. To do this, we move the silver/dielectric sphere on the *xoz*-plane from point to point and calculated the optical force exerted on the sphere. The wavelength of the incident plane wave is fixed at $\lambda = 530$ nm . Figure 15 shows the numerical results given by the FMBEM, where (a) and (b) are the results of the silver sphere while (c) and (d) are the results of the dielectric sphere. Figure 15(a) and Figure 15(c) show that, for both the dielectric and the silver spheres, when one move the sphere along the center axis of the torus, the optical force encounters a zero point where the equilibrium can be achieved. The Fano-shape resonance is due to the symmetry of the structure. The force approaches the photon pressure when the sphere and the torus are infinitely separated and when the sphere is located at the mass center of the torus. The force reaches a maximum at a certain distance $\Delta z$ . Figure 15(b) and Figure 15(d) show the optical forces exerted on the sphere when the sphere travels along *x*-axis. In this case the separation along *z*-axis between the sphere and the torus is fixed at $\Delta z = 50$ nm . We note that for both the silver case and the dielectric case $F_x$ encounters three zero points in the range considered. The zero points at $\Delta z = \pm 40$ nm are two equilibrium points which indicate trapping while the zero point at $\Delta z = 0$ nm is a non-equilibrium point. Since $\Delta z = \pm 40$ nm means the sphere is approximately located at the torus boundary, where the resonance fields are the strongest, this trapping force is mainly a gradient force [Kang *et al.*, 2012]. We conclude from above discussions that the silver torus structure can trap a metallic/dielectric particle at the torus' boundary.

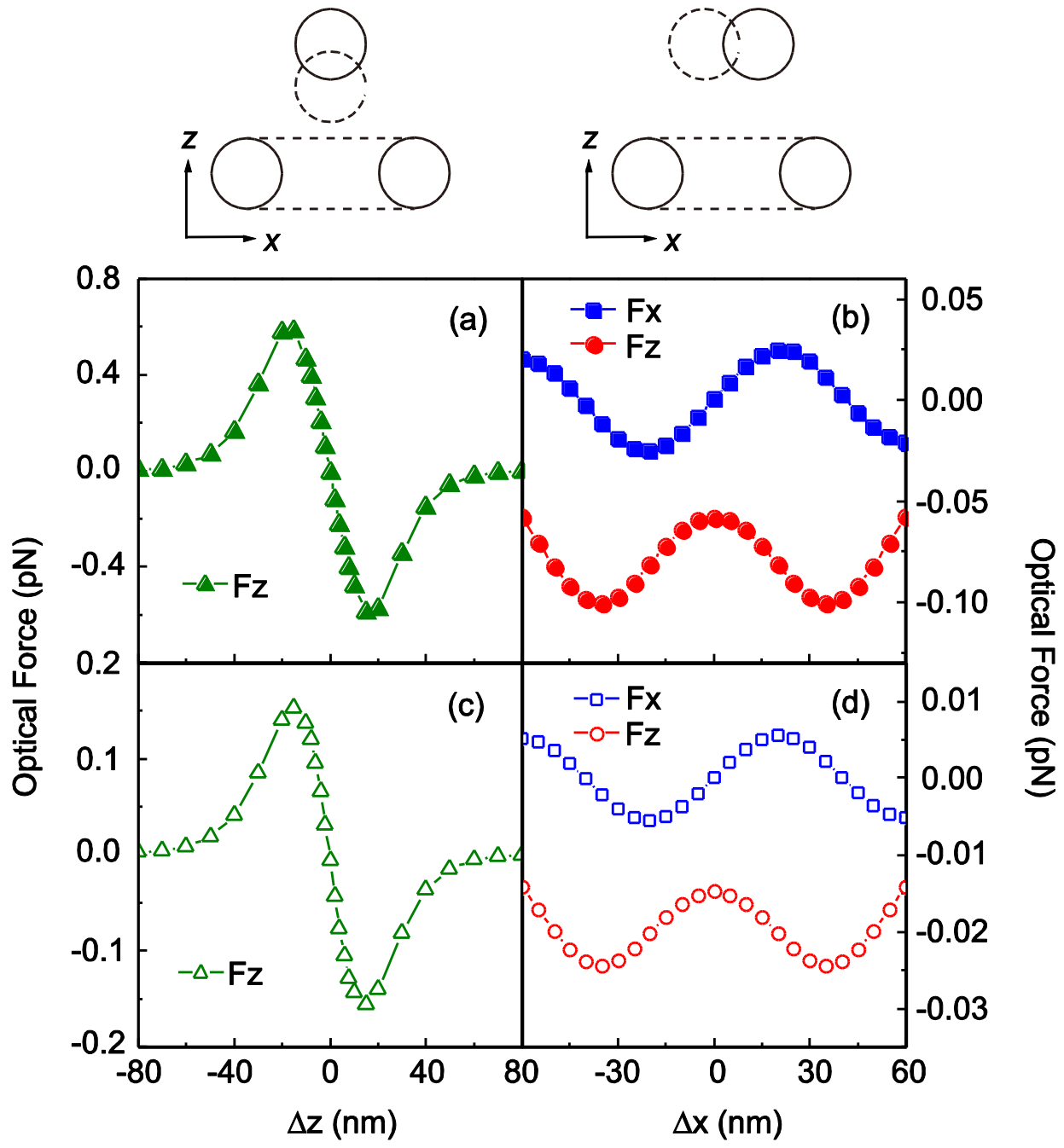

Fig. 15. Optical force between the silver/dielectric sphere and the silver torus as a function of their relative positions. (a)(c) Optical force along $z$-axis exerted on the silver and dielectric sphere, respectively. (b)(d) Optical forces along $x$-axis and $z$-axis exerted on the silver and dielectric spheres, respectively. $(\Delta x, \Delta y, \Delta z)$ denotes the differences between the Cartesian coordinates of the mass centers of the sphere and the torus, where in all the cases considered $\Delta y = 0$ . In (a) and (c) the sphere moves along z-axis while in (b) and (d) the sphere moves along $x$-axis.

## 5. Conclusion

We developed a FMBEM based on boundary integral formulations and Nyström's quadrature rules. The method employs the iterative fGMRES to solve a set of discretized linear equations. The matrix-vector multiplications during iterations are accelerated by the FMM resulting in an $O(n_{\text{iter}} N \log N)$ method. A block-diagonal pre-conditioner is utilized to improve the convergence of the method. We showed using various examples that our method is applicable to scattering problems with frequencies that range from microwave to optical and material properties that range from dielectric to metallic. By comparing the results of the FMBEM with that of the Mie scattering theory we show that the developed method can give accurate results. We then apply the method to study the plasmonic properties of a torus and a SRR structure in the visible range. In this frequency range, the physical phenomena in the considered structures is dominated by electric resonances. Even in the SRR structure where magnetic resonances are usually considered, the electric resonances take the dominate role in the visible range. We also studied the optical force induced in a sphere-torus structure. We found that a metallic sphere can either be repelled or attracted by a silver torus, depending on the excitation frequency of the incident plane wave. However, a dielectric sphere can only be attracted by the silver torus in the frequency range considered. Light can induce the silver torus to attract and bind the spheres to its boundary and this may provide a useful way to trap colloidal particles optically.

**Acknowledgements**

This work is supported by Hong Kong RGC Collaborative Research Grants HKUST2/CRF/11G and CUHK1/CRF/12G and HKUST internal grant SRFI11SC07. S. B. Wang's studentship is supported by Nano Science and Technology Program, Hong Kong University of Science and Technology.

**Appendix A. Derivation of the electric field boundary integral equations**

The Helmholtz equation and Green's function equation corresponding to an arbitrary domain $\Omega$ ( $\Omega = \Omega_i$ and we omit the subscript in the following derivations for simplicity) are expressed as

$$\nabla' \times \left[ \nabla' \times \mathbf{E}(\mathbf{r}') \right] - k^2 \mathbf{E}(\mathbf{r}') = i\omega\mu \mathbf{j}(\mathbf{r}'), \tag{A.1}$$

$$\nabla' \times \left[ \nabla' \times \bar{\mathbf{G}}(\mathbf{r}',\mathbf{r}) \right] - k^2 \bar{\mathbf{G}}(\mathbf{r}',\mathbf{r}) = \delta(\mathbf{r}' - \mathbf{r}) \bar{\mathbf{I}}, \tag{A.2}$$

where $\mathbf{r}' \in \Omega$ and the evaluation point $\mathbf{r}$ is arbitrarily located. The operator $\nabla'$ denotes the partial differentiation with respect to $\mathbf{r}'$. We multiply Eq. (A.2) with $\mathbf{E}(\mathbf{r}')$ from the left and Eq. (A.1) with $\bar{\mathbf{G}}(\mathbf{r}',\mathbf{r})$ from the right, and then subtract the two equations to obtain

$$\mathbf{E}(\mathbf{r}') \cdot \left[ \nabla' \times (\nabla' \times \bar{\mathbf{G}}) \right] - \left[ \nabla' \times \left( \nabla' \times \mathbf{E}(\mathbf{r}') \right) \right] \cdot \bar{\mathbf{G}} = \mathbf{E}(\mathbf{r}') \delta(\mathbf{r}' - \mathbf{r}) - i\omega\mu \mathbf{j}(\mathbf{r}') \cdot \bar{\mathbf{G}} \tag{A.3}$$

We then take the volume integral of the above equation over domain $\Omega$. The right-hand side of Eq. (A.3) becomes

$$\int_\Omega \mathrm{d}V' \left[ \mathbf{E}(\mathbf{r}') \delta(\mathbf{r}' - \mathbf{r}) - i\omega\mu \mathbf{j}(\mathbf{r}') \cdot \bar{\mathbf{G}} \right] = \begin{cases} \mathbf{E}(\mathbf{r}) - \mathbf{E}^{\mathrm{inc}}(\mathbf{r}) & \mathbf{r} \in \Omega \\ -\mathbf{E}^{\mathrm{inc}}(\mathbf{r}) & \mathbf{r} \notin \Omega \end{cases}. \tag{A.4}$$

If we consider $\mathbf{r}$ located inside the domain $\Omega$, the volume integral on the left-hand side of Eq. (A.3) has to be addressed carefully as it is hyper-singular at $\mathbf{r} = \mathbf{r}'$. We extract this singularity by excluding an infinitesimal spherical volume $\odot$ centered at $\mathbf{r}$ [Mittra, 1987], the remaining integral becomes

$$\begin{aligned}
&\int_{\Omega-\odot} \mathrm{d}V' \left\{ \mathbf{E}(\mathbf{r}') \cdot \left[ \nabla' \times (\nabla' \times \bar{\mathbf{G}}) \right] - \left[ \nabla' \times \left( \nabla' \times \mathbf{E}(\mathbf{r}') \right) \right] \cdot \bar{\mathbf{G}} \right\} \\
&= \int_{\Omega-\odot} \mathrm{d}V' \left\{ \begin{aligned} &\mathbf{E}(\mathbf{r}') \cdot \left[ \nabla'(\nabla' \cdot \bar{\mathbf{G}}) \right] - \mathbf{E}(\mathbf{r}') \cdot \nabla'^2 \bar{\mathbf{G}} \\ &- \left[ \nabla' \left( \nabla' \cdot \mathbf{E}(\mathbf{r}') \right) \right] \cdot \bar{\mathbf{G}} + \nabla'^2 \mathbf{E}(\mathbf{r}') \cdot \bar{\mathbf{G}} \end{aligned} \right\} \\
&= \int_{\Omega-\odot} \mathrm{d}V' \left\{ \mathbf{E}(\mathbf{r}') \cdot \left[ \nabla'(\nabla' \cdot \bar{\mathbf{G}}) \right] - \mathbf{E}(\mathbf{r}') \cdot \nabla'^2 \bar{\mathbf{G}} + \nabla'^2 \mathbf{E}(\mathbf{r}') \cdot \bar{\mathbf{G}} \right\}.
\end{aligned} \tag{A.5}$$

The integral of $[\nabla'(\nabla' \cdot \mathbf{E}(\mathbf{r}'))] \cdot \bar{\mathbf{G}}$ vanishes since we only consider localized sources in domain $\Omega$ .The right-hand-side terms of Eq. (A.5) can be transformed into the following forms by invoking Green's second identity,

$$\begin{aligned}
&\int_{\Omega-\odot} \mathrm{d}V' \left[ \mathbf{E}(\mathbf{r}') \cdot \nabla'(\nabla' \cdot \bar{\mathbf{G}}) \right] \\
&= \oint_{\partial\Omega+\partial\odot} \mathrm{d}A' \left\{ \mathbf{E}(\mathbf{s}') \cdot \nabla' \left( \hat{n}' \cdot \bar{\mathbf{G}} \right) - \nabla' \left[ \hat{n}' \cdot \mathbf{E}(\mathbf{s}') \right] \cdot \bar{\mathbf{G}} \right\},
\end{aligned} \tag{A.6}$$

$$\begin{aligned}
&\int_{\Omega-\odot} \mathrm{d}V' \left[ \nabla'^2 \mathbf{E}(\mathbf{r}') \cdot \bar{\mathbf{G}} - \mathbf{E}(\mathbf{r}') \cdot \nabla'^2 \bar{\mathbf{G}} \right] \\
&= \oint_{\partial\Omega+\partial\odot} \mathrm{d}A' \left\{ \left[ \hat{n}' \cdot \nabla' \mathbf{E}(\mathbf{s}') \right] \cdot \bar{\mathbf{G}} - \mathbf{E}(\mathbf{s}') \cdot \left[ \hat{n}' \cdot \nabla' \bar{\mathbf{G}} \right] \right\},
\end{aligned} \tag{A.7}$$

where $\partial\Omega$ denotes the boundary of domain $\Omega$ while $\partial\odot$ denotes the boundary of the infinitesimal sphere $\odot$ . $\hat{n}'$ is the unit normal vector pointing toward outside of the domain $\Omega$ at position $\mathbf{s}'$ (but it points inside $\odot$ as shown in Fig. 1). Note $\mathbf{r}'$ has been replaced by $\mathbf{s}'$ to indicate that it lies on the boundary. Combine Eqs. (A.3)-(A.7) and after some tedious manipulations we can obtain

$$\oint_{\partial\Omega+\partial\odot} \mathrm{d}A' \left\{ \begin{aligned} &\left[ \hat{n}' \nabla' g - \hat{n}' \cdot \nabla' g \bar{\mathbf{I}} - \nabla' g \hat{n}' \right] \cdot \mathbf{E}(\mathbf{s}') \\ &- g \bar{\mathbf{I}} \cdot \left[ \hat{n}' \times (\nabla' \times \mathbf{E}(\mathbf{s}')) \right] \end{aligned} \right\} = -\mathbf{E}^{\mathrm{inc}}(\mathbf{r}), \quad \mathbf{r} \in \Omega \tag{A.8}$$

where the scalar green's function is defined by Eq. (2.4). Note the right-hand side becomes $-\mathbf{E}^{\mathrm{inc}}(\mathbf{r})$ instead of $\mathbf{E}(\mathbf{r}) - \mathbf{E}^{\mathrm{inc}}(\mathbf{r})$ since we have extracted the singularity. In the next step we approach the evaluation point $\mathbf{r}$ to the boundary $\partial\Omega$ from inside, in which case the infinitesimal sphere $\odot$ will be intersected by $\partial\Omega$. In order to circumvent this situation we mathematically curve $\partial\Omega$ around the sphere as shown in Fig. 1. Let us first calculate the contribution from the integral over $\partial\odot$. Denote the radius of $\odot$ as $\upsilon$ and define

$$f(\upsilon) = -\frac{e^{ik\upsilon}}{4\pi\upsilon^2} + ik\frac{e^{ik\upsilon}}{4\pi\upsilon}. \tag{A.9}$$

The surface integral over $\partial\odot$ can be evaluated as

$$\begin{aligned}
&\lim_{\upsilon\to 0}\oint_{\partial\odot} dA'\begin{Bmatrix}\left[\hat{n}'\nabla' g-\hat{n}'\cdot\nabla' g\overline{\mathbf{I}}-\nabla' g\hat{n}'\right]\cdot\mathbf{E}(\mathbf{s}')\\ -g\overline{\mathbf{I}}\cdot\left[\hat{n}'\times(\nabla'\times\mathbf{E}(\mathbf{s}'))\right]\end{Bmatrix}\\
&=\lim_{\upsilon\to 0}\oint_{\partial\odot} dA'\left[-\hat{n}'\hat{n}' f(\upsilon)+f(\upsilon)\overline{\mathbf{I}}+f(\upsilon)\hat{n}'\hat{n}'\right]\cdot\mathbf{E}(\mathbf{s})\\
&\quad-\lim_{\upsilon\to 0}\oint_{\partial\odot} dA'\left(g\overline{\mathbf{I}}\times\hat{n}'\right)\cdot\left[\nabla'\times\mathbf{E}(\mathbf{s})\right]\\
&=\lim_{\upsilon\to 0}4\pi\upsilon^2 f(\upsilon)\mathbf{E}(\mathbf{s})-0=-\mathbf{E}(\mathbf{s}).
\end{aligned}\tag{A.10}$$

The second integral in Eq. (A.10) vanishes due to rotation symmetry. We can calculate the integral over the spherically curved part of $\partial\Omega$ by the same way and it gives a result of $(\Theta/4\pi)\mathbf{E}(\mathbf{s})$, where $\Theta$ represents the solid angle subtended by this part of boundary. For a locally smooth boundary we have $\Theta=2\pi$. Combining this with Eq. (A.8) and Eq. (A.10), we have

$$\oint_{\partial\Omega}^{\mathrm{CPV}} \mathrm{d}A'\begin{Bmatrix}\left[\hat{n}'\nabla' g-\hat{n}'\cdot\nabla' g\overline{\mathbf{I}}-\nabla' g\hat{n}'\right]\cdot\mathbf{E}(\mathbf{s}')\\ -g\overline{\mathbf{I}}\cdot\left[\hat{n}'\times(\nabla'\times\mathbf{E}(\mathbf{s}'))\right]\end{Bmatrix}=\frac{1}{2}\mathbf{E}(\mathbf{s})-\mathbf{E}^{\mathrm{inc}}(\mathbf{s}),\tag{A.11}$$

where "CPV" indicates the integral is a Cauchy-Principle-Value integral. The above equation is usually referred to as the electric field integral equation in the literatures. We note the derivations above can be applied to any domain of a scattering system.

## Appendix B. Analytical integration over triangular element

Eq. (2.12) explicitly involves weak ( $g$ ) and strong ( $\nabla g$ ) singularities, which corresponding to the self-interactions. Here we address the integrals in an analytical way. Let us first define (Fig. B.1)

$$\hat{n}=\frac{\overrightarrow{AB}\times\overrightarrow{AC}}{\left|\overrightarrow{AB}\times\overrightarrow{AC}\right|},\ \hat{i}=\frac{\overrightarrow{OD}}{\left|\overrightarrow{OD}\right|},\ \hat{j}=\hat{n}\times\hat{i},\tag{B.1}$$

where $\hat{n}$ is the unit normal vector. $\hat{i}$ and $\hat{j}$ define a local two-dimensional Cartesian coordinates with respect to the triangular element. We employ polar coordinates $(\rho,\phi)$ to do the integral.

For the integral related to $g$ we have

$$\int_{\vartriangle} g\mathrm{d}A=\int_{\vartriangle}\frac{e^{ik\rho}}{4\pi\rho}\rho\mathrm{d}\rho\mathrm{d}\phi\approx\frac{1}{4\pi}\int_{\vartriangle}\mathrm{d}\rho\mathrm{d}\phi+\frac{ikS_{\vartriangle}}{4\pi},\tag{B.2}$$

where $S_{\Delta}$ is the area of the triangle $\Delta ABC$ and we have done the Taylor expansions of the integrand. The integral of the first term in Eq. (B.2) can be expressed as

$$
\begin{aligned}
\int_{\Delta} \mathrm{d}\rho \mathrm{d}\phi &= \int_{-\theta_1}^{\theta_2} \frac{\left|\overrightarrow{OD}\right|}{\cos\phi}\mathrm{d}\phi + \int_{-\theta_3}^{\theta_4} \frac{\left|\overrightarrow{OE}\right|}{\cos\phi}\mathrm{d}\phi + \int_{-\theta_5}^{\theta_6} \frac{\left|\overrightarrow{OF}\right|}{\cos\phi}\mathrm{d}\phi \\
&= |\overrightarrow{OD}| \ln \frac{\tan(\frac{\pi}{4}+\frac{\theta_1}{2})}{\tan(\frac{\pi}{4}-\frac{\theta_2}{2})} + |\overrightarrow{OE}| \ln \frac{\tan(\frac{\pi}{4}+\frac{\theta_3}{2})}{\tan(\frac{\pi}{4}-\frac{\theta_4}{2})} + |\overrightarrow{OF}| \ln \frac{\tan(\frac{\pi}{4}+\frac{\theta_5}{2})}{\tan(\frac{\pi}{4}-\frac{\theta_6}{2})}.
\end{aligned} \tag{B.3}
$$

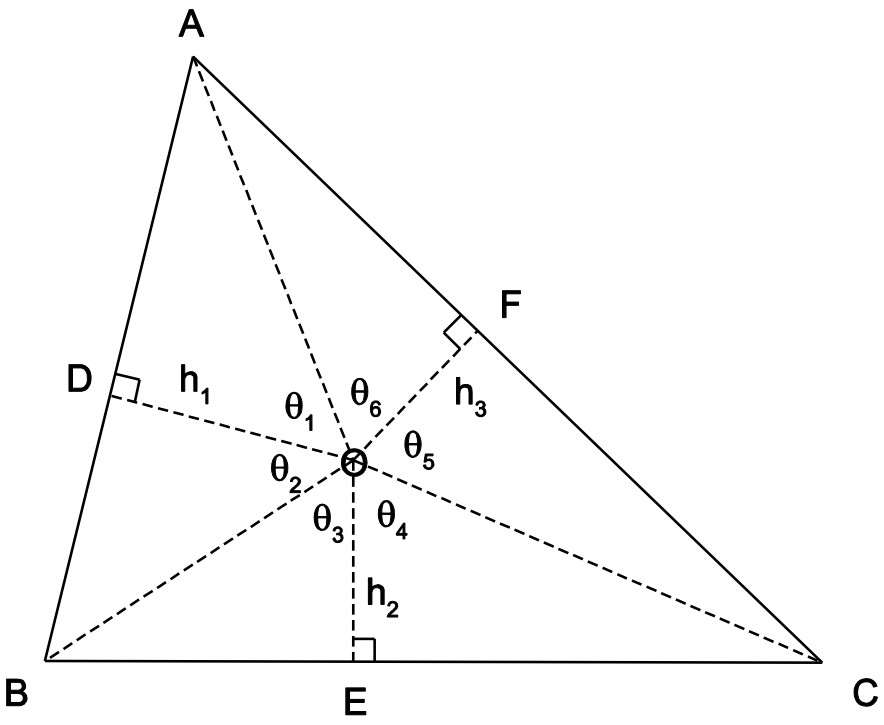

Fig. B.1. Analytical integration on a triangular element in the FMBEM. The unit normal vector $\hat{n}$ is perpendicular to the triangle surface.

For the integral related to $\nabla g$ we have

$$
\begin{aligned}
\int_{\Delta} \nabla g \mathrm{d}A &= \frac{1}{4\pi}\left[\int_{\Delta} \nabla\left(\frac{e^{ikr}-1}{r}\right)\mathrm{d}A - \int_{\Delta}\frac{\mathbf{r}}{r^3}\mathrm{d}A\right] \\
&\approx \frac{1}{4\pi}\left[\int_{\Delta} \nabla\left(ik - \frac{1}{2}k^2 r\right)\mathrm{d}A - \int_{\Delta}\frac{\mathbf{r}}{r^3}\mathrm{d}A\right] \\
&= \frac{-k^2}{8\pi}\int_{\Delta}\frac{\mathbf{r}}{r}\mathrm{d}A - \frac{1}{4\pi}\int_{\Delta}\frac{\mathbf{r}}{r^3}\mathrm{d}A.
\end{aligned} \tag{B.4}
$$

The first integral on the right side of Eq. (B.4) can be done as

$$
\begin{aligned}
\int_{\Delta OAB} \frac{\mathbf{r}}{r}\mathrm{d}A &= \int_{\Delta OAB} \hat{\rho}\rho \mathrm{d}\rho \mathrm{d}\phi \\
&= \frac{\left|\overrightarrow{OD}\right|^2}{2}\int_{-\theta_1}^{\theta_2} \frac{\cos\phi \hat{i} + \sin\phi \hat{j}}{\cos^2\phi}\mathrm{d}\phi \\
&= \frac{|\overrightarrow{\mathrm{OD}}|}{2}\left[\ln\frac{\tan(\frac{\pi}{4}+\frac{\theta_1}{2})}{\tan(\frac{\pi}{4}-\frac{\theta_2}{2})}\overrightarrow{\mathrm{OD}} + \left(\frac{1}{\cos\theta_2} - \frac{1}{\cos\theta_1}\right)\hat{n}\times\overrightarrow{\mathrm{OD}}\right],
\end{aligned} \tag{B.5}
$$

where we focus just on the integration over $\Delta OAB$ . Similar expressions hold for the integral over $\Delta OAC$ and $\Delta OBC$ . The second part of Eq. (B.4) is a CPV integral that can be done as

$$\begin{aligned}
\lim_{\upsilon\to 0}\int_{\Delta OAB-\odot}\frac{\mathbf{r}}{r^3}\mathrm{d}A &= \lim_{\upsilon\to 0}\int_{\Delta OAB-\odot}\frac{\hat{\rho}}{\rho}\mathrm{d}\rho\mathrm{d}\phi \\
&= \lim_{\upsilon\to 0}\int_{-\theta_1}^{\theta_2}\ln\frac{\rho}{\upsilon}\left(\cos\phi\hat{i}+\sin\phi\hat{j}\right)\mathrm{d}\phi \\
&= \left[\lim_{\upsilon\to 0}\left(\ln\frac{\left|\overrightarrow{OD}\right|}{\upsilon}\right)\left(\sin\phi\hat{i}-\cos\phi\hat{j}\right)\right]_{-\theta_1}^{\theta_2} \\
&\quad -\int_{-\theta_1}^{\theta_2}\ln\cos\phi\left(\cos\phi\hat{i}+\sin\phi\hat{j}\right)d\phi,
\end{aligned} \tag{B.6}$$

where we have removed the singular point by extracting an infinitesimal circle of radius $\upsilon$. The first part of Eq. (B.6) can be transformed as

$$\lim_{\upsilon\to 0}\left(\ln\frac{\left|\overrightarrow{OD}\right|}{\upsilon}\right)\left(\sin\phi\hat{i}-\cos\phi\hat{j}\right)\Big|_{-\theta_1}^{\theta_2} = \lim_{\upsilon\to 0}\left(\ln\frac{\left|\overrightarrow{OD}\right|}{\upsilon}\right)\hat{n}\times\left(\frac{\overrightarrow{OA}}{\left|\overrightarrow{OA}\right|}-\frac{\overrightarrow{OB}}{\left|\overrightarrow{OB}\right|}\right), \tag{B.7}$$

while the second part is

$$\begin{aligned}
&\int_{-\theta_1}^{\theta_2}\ln\cos\phi\left(\cos\phi\hat{i}+\sin\phi\hat{j}\right)\mathrm{d}\phi \\
&= \left[\left(\frac{1}{2}\ln\frac{1+\sin\phi}{1-\sin\phi}+\sin\phi\ln\cos\phi-\sin\phi\right)\hat{i}\right]_{-\theta_1}^{\theta_2} \\
&\quad +\left[\left(\cos\phi-\cos\phi\ln\cos\phi\right)\hat{j}\right]_{-\theta_1}^{\theta_2}.
\end{aligned} \tag{B.8}$$

Similarly we can obtain the integral over $\Delta OAC$ and $\Delta OBC$ . Sum up the contributions from all the three parts and we have the final expression written as

$$\int_{\Delta}\frac{\mathbf{r}}{r^3}\mathrm{d}A = \frac{\overrightarrow{OD}}{\left|\overrightarrow{OD}\right|}\ln\frac{\tan\left(\frac{\pi}{4}-\frac{\theta_2}{2}\right)}{\tan\left(\frac{\pi}{4}+\frac{\theta_1}{2}\right)}+\frac{\overrightarrow{OE}}{\left|\overrightarrow{OE}\right|}\ln\frac{\tan\left(\frac{\pi}{4}-\frac{\theta_4}{2}\right)}{\tan\left(\frac{\pi}{4}+\frac{\theta_3}{2}\right)}+\frac{\overrightarrow{OF}}{\left|\overrightarrow{OF}\right|}\ln\frac{\tan\left(\frac{\pi}{4}-\frac{\theta_6}{2}\right)}{\tan\left(\frac{\pi}{4}+\frac{\theta_5}{2}\right)}. \tag{B.9}$$

## Appendix C. The octree data structure and box relationships in the FMBEM

The FMBEM employs a hierarchical octree data structure that guarantees an efficient implementation of the matrix-vector multiplications. To construct such a data structure, we consider a sphere as shown in Fig. C.1. The discretized sphere is first enclosed by a cubic box on level 0. On level 1, the box is subdivided into eight identical smaller boxes with each one containing a few elements. Then on level 2 each of the smaller boxes is further subdivided into eight boxes. This procedure is repeated until the smallest box contains sufficient few elements. Each box on a level can be identified with a global index ($l$,$q$) with $l$ being the level number and $q$ being the $q$th box on this level. The empty boxes on each level (boxes do not contain elements) are ruled out in the FMBEM.

With the octree data structure, we make the following definitions. For the boxes that are contained in the same box ($l$,$q$), they are referred to as the children boxes of box ($l$,$q$) and denoted as "Children($l$,$q$)". Similarly, "Parent($l$,$q$)" denotes the parent box of box ($l$,$q$). Figure C.2 shows a case of S|S translation applied to the "Children($l$,$q$)" in the left-up corner. For the boxes that have common corners/edges/faces with box ($l$,$q$), they are referred to as the neighbor boxes of box ($l$,$q$) and denoted as "Neighbor($l$,$q$)". There is also a kind of boxes denoted as "FF($l$,$q$)", which are the far-field boxes with respect to box ($l$,$q$). The "FF($l$,$q$)" boxes are chosen based on the condition: $\mathrm{FF}(l,q) \in \mathrm{Children}(\mathrm{Neighbor}(\mathrm{Parent}(l,q)))$ and $\mathrm{FF}(l,q) \notin \mathrm{Neighbor}(l,q)$ , which guarantees the validity of the far-field expansions[Gumerov and Duraiswami, 2004]. Figure C.2 shows a case of S|R translation applied to the "FF($l$,$q$)" boxes in the right-down corner.

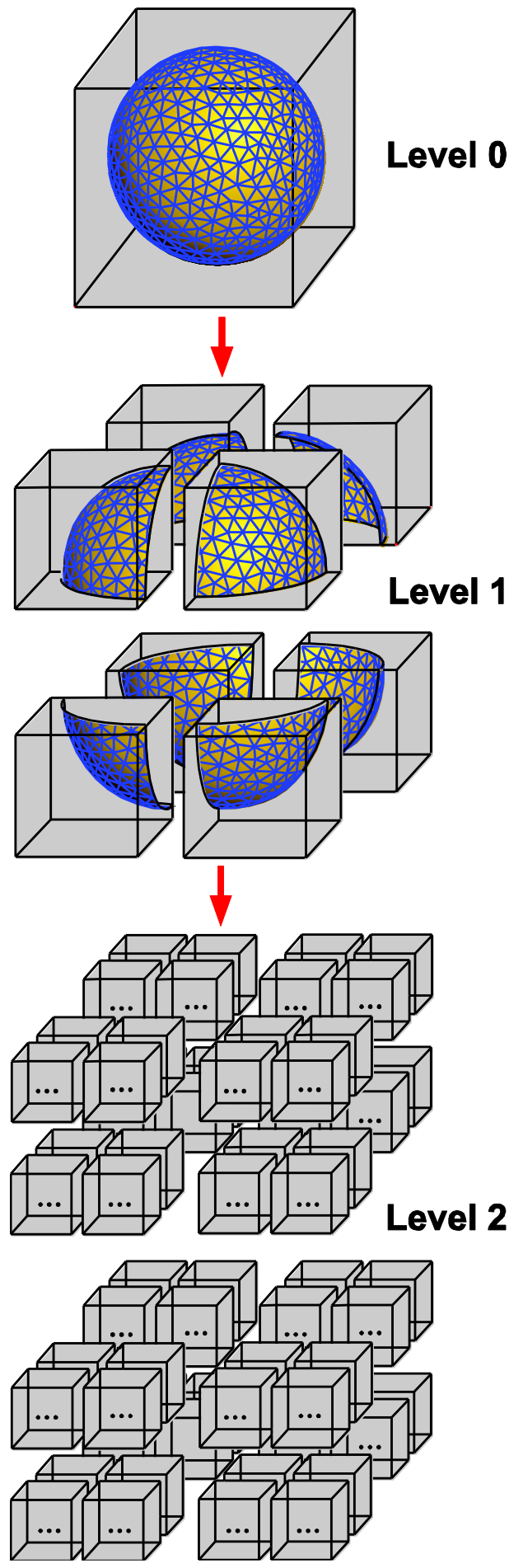

Fig. C.1. Hierarchical octree data structure in the FMBEM.

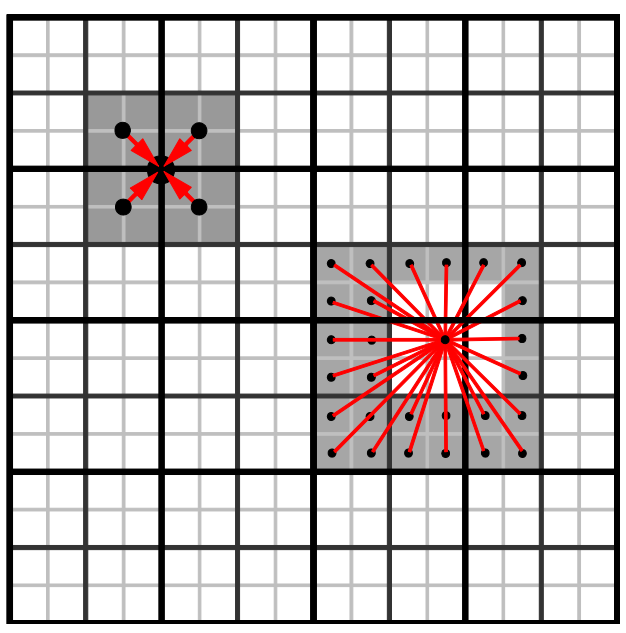

Fig. C.2. Translation schemes in the FMBEM. On the left-up corner we show a case of S|S translation from the children boxes to the parent box. The right-down corner we show a case of S|R translations applied to the $\mathrm{FF}(l,q)$ boxes (the dark boxes).

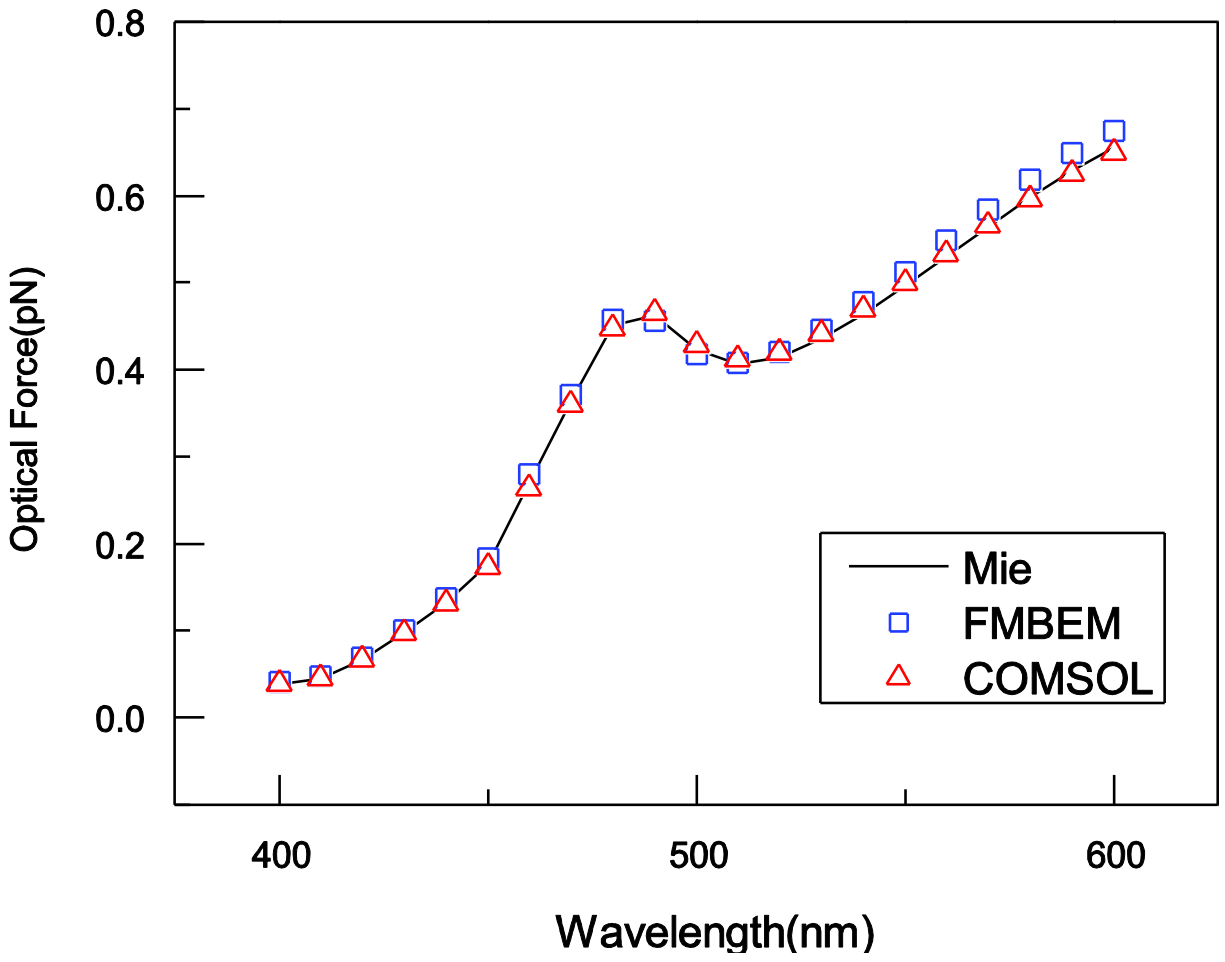

Fig. D.1 Optical force acting on the left sphere calculated by Mie theory, FMBEM and the commerical package COMSOL.

Table D.1 Solution information of the COMSOL and the FMBEM

| Method | No. of unknowns | Memory | Solution time | Error |
| --- | --- | --- | --- | --- |
| COMSOL | $3.1\times10^5$ | 8Gb | 1h 36min | 0.01 |
| FMBEM | $2.0\times10^4$ | 400Mb | 6min | 0.03 |

## Appendix D. Numerical performance of the FMBEM on calculating optical forces

Here we provide an prototypical example of applying the FMBEM to calculate optical forces and at the same time we compare it to the results of the analytical Mie theory and a commercial finite element method package COMSOL. We again consider the double-gold-sphere case introduced in Section 3.2 and the configuration is shown by the inset of the first panel in Fig. 6. We calculated the time-averaged optical forces acted on the left sphere by applying Maxwell stress tensor approach. This is done by integrating the time-averaged Maxwell stress tensor $\langle\overline{\mathbf{T}}\rangle = 1/2[\varepsilon_0\mathbf{E}\mathbf{E}^* + \mu_0\mathbf{H}\mathbf{H}^* - 1/2(\varepsilon_0\mathbf{E}\cdot\mathbf{E}^* + \mu_0\mathbf{H}\cdot\mathbf{H}^*)]$ over a closed surface surrounding the sphere. Figure D.1 shows the results obtained by the three kinds of methods, where we plot the $z$-component time-averaged optical force as a function of the wavelength. We took without loss of generality the intensity of the incident plane wave to be $1\,\mathrm{mW}/(\mu\mathrm{m})^2$. We also provide a summary of relevant information of the two numerical methods (COMSOL and FMBEM) on addressing this particular example in Table D.1. The FMBEM has a total complexity of $O(N\log N)$ while the finite element package COMSOL has a nominal complexity of $O(N)$. However, the total number of unknowns $N$ of the COMSOL is much larger ( $N = 3.1\times10^5$ ) than that of the FMBEM ( $N = 2\times10^4$ ) since the unknowns of the FMBEM only lie on the boundaries. We tested the two methods on a single CPU of the same computer. By both using an iterative solver, the COMSOL had a 8Gb memory consumption and took 1 hour and 36 minutes to get a converged solution (residue error = 0.001) for a single sampling point in Fig. D.1. However, the FMBEM only need 400Mb memory and 6 minutes to get the solution (with the same residue error). We compared the solutions obtained by the two methods with the Mie theory results by defining the solution error as: $error = \sum_{i=1}^{n} | F_{\text{Mie}} - F_x | / \sum_{i=1}^{n} | F_{\text{Mie}} |$, where $F_{\text{Mie}}$ is the force results given by the Mie theory and $F_x$ denotes the results of the COMSOL/FMBEM; $n$ is the total number of wavelength sampling points in Fig. D.1. We note the error of the FMBEM ( $error = 0.03$ ) is comparable to that of the COMSOL( $error = 0.01$ ). But considering the highly advantage of much less memory and time consumptions of the FMBEM, the developed method should be a good tool for addressing optical force problems associated to plasmonic systems.

* Corresponding author